\documentclass[12pt]{article}

\usepackage{graphicx}
\usepackage{amsmath}
\usepackage{amssymb}
\usepackage{amsthm}
\usepackage{xspace}
\usepackage{geometry}
\usepackage{color}

\input epsf

\def\appendix#1{
  \addtocounter{section}{1} 
 \setcounter{equation}{0}
  \renewcommand{\thesection}{\Alph{section}}
 \section*{Appendix \thesection\protect\indent \parbox[t]{11.715cm} {#1}}
  \addcontentsline{toc}{section}{Appendix \thesection\ \ \ #1}
  }

\renewcommand{\thefootnote}{\fnsymbol{footnote}}

\numberwithin{equation}{section}

\begin{document}

\newcommand{\fuud}[3]{
f{_{\hspace{-6pt}#1#2}\atop ^{\hspace{9pt} #3} }
}
\newcommand{\fudd}[3]{
f{_{\hspace{-9pt}#1}\atop ^{\hspace{4pt} #2#3} }
}
\newcommand{\fduu}[3]{
f{_{\hspace{4pt}#2#3}\atop ^{\hspace{-15pt} #1} }
}
\newcommand{\fddu}[3]{
f{_{\hspace{8pt}#3}\atop ^{\hspace{-8pt} #1#2} }
}

\newcommand{\llra}{ {\bf \longleftrightarrow} }
\newcommand{\lra}{ {\bf \longrightarrow} }
\newcommand{\lla}{ {\bf \longleftarrow} }
\newcommand{\psu}{\mathfrak{psu}}
\newcommand{\su}{\mathfrak{su}}
\newcommand{\lbr}{[ \! [}
\newcommand{\rbr}{] \! ]}
\newcommand{\slot}{\hspace{0.4pt}\lfloor\hspace{-3pt}\rceil\hspace{-0.4pt}} 

\newcommand{\spacemaker}[1]{
\setbox1=\hbox{$#1$}
\setbox2=\hbox to \wd1 {}
\box2
}


\thispagestyle{empty}
\begin{flushright}\footnotesize
\texttt{arxiv:0706.1525}\\
\texttt{CALT-68-2649}\\
\end{flushright}

\renewcommand{\thefootnote}{\fnsymbol{footnote}}
\setcounter{footnote}{0}
\vspace{1.5cm}

\begin{center}
{\Large\textbf{
Perturbative study of the transfer matrix \\ 
on the string worldsheet in $AdS_5 \times S^5$ 
}\par}

\vspace{1.5cm}

\textrm{Andrei Mikhailov and Sakura Sch\"afer-Nameki}
\vspace{1.2cm}

\textit{California Institute of Technology\\
1200 E California Blvd., Pasadena, CA 91125, USA } \\
\texttt{andrei@theory.caltech.edu, ss299@theory.caltech.edu} \vspace{3mm}


\par\vspace{1.2cm}

\textbf{Abstract}\vspace{5mm}
\end{center}

\noindent 
Quantum non-local charges are central to  the quantum integrability 
of a sigma-model. 
In this paper we study the quantum consistency and UV finiteness 
of non-local charges of 
string theory in $AdS_5 \times S^5$. We use the pure spinor formalism.
We develop the near-flat space expansion of the transfer 
matrix and calculate the one-loop divergences. We find that the
logarithmic divergences cancel at the level of one loop. 
This gives strong support 
to the quantum integrability of the full string theory. 
We develop a calculational setup for the renormalization group analysis of 
Wilson line type of operators on the string worldsheet.

\vspace*{\fill}

\setcounter{page}{1}
\renewcommand{\thefootnote}{\arabic{footnote}}
\setcounter{footnote}{0}

\newpage

\tableofcontents


\section{Introduction}

The integrability of $\mathcal{N}=4$ SYM and string theory on 
$AdS_5 \times S^5$ has emerged as a major tool in testing the AdS/CFT 
correspondence. 
Despite the impressive progress that the assumption of integrability has 
enabled, relatively little is known about the integrable structure that 
underlies both theories. Understanding this would be desirable for a 
number of reasons: the showpiece of AdS/CFT at present is undoubtedly the 
asymptotic S-matrix \cite{Beisert:2006ez}, based on the earlier works 
\cite{Arutyunov:2004vx,  Arutyunov:2006iu, Beisert:2005tm, Eden:2006rx, Beisert:2006ib},
which seems to correctly capture the scattering in infinite volume 
for both strings and SYM. Applicability of factorized scattering assumes
 the existence of quantum non-local 
conserved charges \cite{Zamolodchikov:1978xm, Luscher:1977uq}. 
Furthermore, the S-matrix alone does not describe the full spectrum, 
as it assumes scattering in infinite volume and thus fails to capture 
finite-size corrections. This shortcoming was made explicit in the string 
theory in \cite{Schafer-Nameki:2006ey} and for $\mathcal{N}=4$ SYM 
in \cite{Kotikov:2007cy}. A possible way to fix this problem is to apply 
the thermodynamic Bethe-ansatz procedure, 
as suggested in \cite{Ambjorn:2005wa}, which however may be limited to 
very specific, low-lying states. Alternatively, a systematic procedure, 
based on the Baxter $Q$-operator, was proposed in  
\cite{Bytsko:2006ut, Teschner:2007ng}, which  in the case of sinh-Gordon 
theory enables to compute the finite-size effects for all states. 
Key to this analysis is however the knowledge of the quantum symmetries 
of the problem -- which in the case of AdS/CFT remains to be uncovered. 

In this paper we would like to take a step in the direction 
of a better understanding of the quantum transfer matrix and the 
associated non-local charges of string theory in $AdS_5 \times S^5$.
We will use the pure spinor formalism. 
One of our motivations is to see 
how far the computational feasability of the pure spinors 
extends beyond flat-space. 
Furthermore, the pure spinor string in 
$AdS_5 \times S^5$ \cite{Berkovits:2000fe, 
Berkovits:2000yr, Berkovits:2001ue, Berkovits:2007zk} has 
various features that make it a 
natural framework for quantum computations. 
The theory is conformal on the world-sheet 
\cite{Berkovits:2004xu,Vallilo:2002mh} and, 
as we shall see, quantum computations can be performed without 
choosing a specific gauge that 
breaks the global symmetries. 
In view of integrability, an interesting combination of integrable 
and conformal structure of 
the world-sheet theory emerges. The existence of classical local 
conserved charges was established 
in \cite{Vallilo:2003nx}. Integrability of the quantum theory was 
anticipated in 
\cite{Berkovits:2004jw, Berkovits:2004xu}, where an argument 
showing BRST-invariance of the non-local 
charges was put forward. 

In this paper we examine the UV finiteness of the transfer matrix.
We analyse the short-distance 
singularities of the currents. 
We perform a perturbative expansion 
around flat space, determine the OPE of the currents 
to the leading order in curvature corrections and
study the logarithmic divergences of the path ordered integrals
of currents.
We find that the logarithmic divergences of the transfer matrix
cancel at the level of one loop. 
This gives support to the claim that the quantum string in 
$AdS_5 \times S^5$ is quantum integrable. 

Important related works in the WZW literature are
\cite{Bachas:2004sy, Alekseev:2007in}. Those papers 
discussed the short distance singularities of loop operators 
in boundary and bulk WZW models. 

The plan of this paper is as follows: in Section \ref{sec:BriefIntroduction}
 we begin with a very brief summary of the pure spinor 
string in $AdS_5 \times S^5$, 
discussing the zero curvature formulation of the equations of motion, and 
the classical transfer matrix. 
In Section \ref{sec:ActionAndOPE} we discuss  the near flat space 
expansion of the action. We study the short distance singularities in the
OPE of the currents and calculate the field renormalization, which is necessary
in our formalism. 
Flat space limit was recently discussed in 
\cite{Berkovits:2007zk}. 
Section \ref{sec:RenormalizationWilsonLine} contains the general discussion of 
the renormalization of the Wilson line type of operators.
In Sections \ref{sec:CalculationOfLogDiv} and \ref{sec:LogDivAndGlobalCharges}
 we calculate the logarithmic divergences of the transfer matrix at the one loop
level, and find that they cancel. 
As a consistency check of our 
calculational framework, we show 
explicitly in Section \ref{sec:InfiniteLine}
 that logarithmic divergences cancel in the global symmetry charge. 
Section \ref{sec:SummaryAndConclusions} is a conclusion.  
Appendix A provides details on the 
algebra $\mathfrak{psu}(2,2|4)$. In Appendix B we
discuss linear divergences; 
they would be important in the higher loop calculations.

\section{Brief introduction to pure spinors}
\label{sec:BriefIntroduction}

\subsection{Action functional and ``capital currents" $J_{\pm}$}
The target space $AdS_5 \times S^5$ is the coset space
\[
{PSU(2,2|4) \over SO(4,1) \times SO(5)} \,.
\]
A point of this space can be parametrized by the group element
$g\in PSU(2,2|4)$ modulo the left shift by $h\in SO(4,1) \times SO(5)$:
\begin{equation}\label{GaugeTransformationCoset}
g\equiv hg\;,\;\;\;
h\in SO(4,1) \times SO(5) \,.
\end{equation}
The action in the pure spinor 
formalism 
\cite{Berkovits:2000fe, Berkovits:2000yr, Berkovits:2001ue, Berkovits:2007zk} 
is constructed out of the ``capital" currents\footnote{We call these currents
``capital" because there are also ``small case" currents; 
see Section \ref{sec:SmallCaseCurrents}. Capital currents are
invariant under the global symmetry $PSU(2,2|4)$. Small case currents
are invariant under the gauge transformations (\ref{GaugeTransformationCoset}).}
\begin{equation}
J_{\pm} = -\partial_{\pm} g g^{-1} \,, 
\qquad g\in {PSU(2,2|4) \over SO(4,1) \times SO(5)}\,,
\end{equation}
and bosonic ghosts $(\lambda^\alpha, w_{\pm\alpha})$. 
The Lie-algebra 
$ \mathfrak{psu}(2,2|4)$ has a $\mathbb{Z}_4$ grading 
\begin{equation}
\mathfrak{psu} (2,2|4) = \mathfrak{g} =
\mathfrak{g}_{\bar{0}} \oplus\mathfrak{g}_{\bar{1}} \oplus \mathfrak{g}_{\bar{2}} \oplus \mathfrak{g}_{\bar{3}} \,.
\end{equation}
This $\mathbb{Z}_4$ grading has a clear physical meaning.
It depends on the choice of a point $x_0$ in $AdS_5\times S^5$. 
This same point will be used in our flat space expansion. The flat
space limit is the limit when the string is localized near this point $x_0$.
In this limit the target space $AdS_5\times S^5$ is approximated by flat
space, which is the tangent space $T_{x_0}(AdS_5\times S^5) $.
 In the flat space Type IIB superstring
there are two supersymmetries, $\epsilon_L$ and $\epsilon_R$, which are
both Majorana-Weyl spinors. One comes from the left sector on the string worldsheet
and the other from the right sector, so they are physically distinguished.
We say that the Killing spinor on $AdS_5\times S^5$ is in
$\mathfrak{g}_{\bar{1}}$ if it becomes $\epsilon_L$ in the flat limit near $x_0$,
and in $\mathfrak{g}_{\bar{3}}$ if it becomes $\epsilon_R$.
This is a grading, because the anticommutator
of two left supersymmetries gives us a translation, while the anticommutator
of left and right supersymmetry gives zero in the flat space limit. 

It is useful to label the generators of 
$\mathfrak{g}=\mathfrak{psu} (2,2|4)$ using this flat
space picture. The bosonic generators in the flat space limit 
are boosts and rotations $t^0_{[\mu\nu]}$, and also the translations
$t^2_{\mu}$; here $\mu$ and $\nu$ are the vector indices of the tangent space.
The fermionic generators are the left supersymmetries $t^3_{\alpha}$ and
the right supersymmetries $t^1_{\dot \alpha}$. Here $\alpha$ and $\dot{\alpha}$
are both the spinor indices of the Majorana-Weyl spinors; they have the same
chirality (left Majorana-Weyl spinors). In other words, 
$\alpha$ vs. $\dot\alpha$ do not indicate different 
chiralities but the grading (3 and 1, respectively). To summarize, we have 
the following set of generators:
\begin{equation}
t = \{t^0_{[\mu\nu]}\,, \ t^1_{\dot \alpha}   \,,\  t^2_{\mu}\,,\ t^3_{\alpha} \,\}\,, 
\end{equation}
where $\mu = 0,\cdots, 9$ and $\alpha, {\dot \alpha}= 1, \cdots 16$.
The bosonic subalgebra is $\mathfrak{g}_{\bar{0}} \oplus \mathfrak{g}_{\bar{2}}$, where 
$\mathfrak{g}_{\bar{0}}$ corresponds to the denominator algebra 
$\mathfrak{so}(4,1) \oplus \mathfrak{so}(5)$. 

The bosonic ghosts 
$(\lambda_3, w_{1+})$ and $(\lambda_1, w_{3-})$ 
take values in $\mathfrak{g}_{\bar{1}} \oplus \mathfrak{g}_{\bar{3}}$ and satisfy the pure spinor condition
\begin{equation}\label{PureSpinorConst}
\lambda_1 \Gamma^\mu \lambda_1 = \lambda_3 \Gamma^\mu \lambda_3 =0 \,, 
\end{equation}
where $\Gamma^\mu_{\alpha \beta}$ are the $SO(9,1)$ gamma-matrices.
The solution space to the pure spinor constraint is eleven complex 
dimensional and is parametrized by  the coset space $SO(10)/U(5)$. 

The action functional is \cite{Berkovits:1999zq, Berkovits:2000fe}
\begin{eqnarray}
S =&&   {R^2\over \pi} \int d^2 z\, \hbox{Str} \Big( {1\over 2} J_{2+}J_{2-} +
{3\over 4} J_{1+}J_{3-} +{1\over 4} J_{3+}J_{1-}  \nonumber \\[5pt]
&&   \qquad      +
w_{1+}\partial_-\lambda_3 + w_{3-}\partial_+\lambda_1 +N_{0+}J_{0-}
+N_{0-}J_{0+}-N_{0+}N_{0-} \Big)  \,,
\end{eqnarray}
where the ghost currents
\begin{equation}
N_{0+}=-\{w_{1+},\lambda_3\} \,,\qquad N_{0-}=-\{w_{3-},\lambda_1\} \,,
\end{equation}
can be seen to couple non-trivially to the physical fields.
The currents are contracted with the generators of $\mathfrak{psu}(2,2|4)$
\begin{equation}
J_{0+} = J_{0+}^{[\mu\nu]} t^0_{[\mu\nu]}\;,\;\;\;
J_{2+} = J_{2+}^{\mu} t^2_{\mu}\;,\;\;\;
J_{3+} = J_{3+}^{\alpha} t^3_{\alpha}\;,\;\;\;
J_{1+} = J_{1+}^{\dot{\alpha}} t^1_{\dot{\alpha}}\,,
\end{equation}
which are chosen in a finite-dimensional representation. 
The ``Str" in the action is the supertrace in the
fundamental representation $4|4$ of $\mathfrak{su}(4|4)$; it defines
the invariant bilinear form on $\mathfrak{psu}(2,2|4)$. 
The physical spectrum is obtained as the cohomology of the BRST-operator. 
The classical BRST-transformation is generated by
\begin{equation}
Q = \int   \, \hbox{Str} 
\left( \lambda_1 J_{3-}d\tau^- + \lambda_3 J_{1+}d\tau^+ \right) \,.
\end{equation}
Using the pure spinor constraint (\ref{PureSpinorConst}) it follows that 
$Q$ is nilpotent up to a gauge 
transformation.   
Quantum BRST and conformal invariance of the action were established in 
\cite{Vallilo:2002mh,Berkovits:2004xu}.


It is probably not very easy to
define the classical limit, due to the pure spinor ghosts;
it is not obvious that just setting $\lambda=w =0$
would yield the correct classical theory. But the main point in
favour of the pure spinor approach is that it is possible to do perturbative
calculations in the quantum theory without introducing the light cone
gauge, and thereby maintaining
the full $\mathfrak{psu}(2,2|4)$ symmetry.

The only serious obstacle in this direction is lack of experience
with the curved $\beta\gamma$ systems; for some recent progress
in this direction see \cite{Nekrasov:2005wg} and references therein.

\subsection{Equations of motion and zero curvature equations}

Classical integrability was established first for the GSMT action 
\cite{Metsaev:1998it} by rewriting the equations of motion in terms of zero-curvature 
equations \cite{Bena:2003wd}. Classical integrability for the pure-spinor string 
can be established likewise. In the following we will review the analysis in \cite{Vallilo:2003nx, Berkovits:2004jw, Berkovits:2004xu}\footnote{Further discussion of the classical dynamics have appeared in \cite{Bianchi:2006im, Kluson:2006wq} }.

The zero curvature conditions on the currents $J$ are
\begin{equation}\label{ZeroCurvature}
        \partial_+J_--\partial_-J_++[J_+,J_-]=0 \,.
\end{equation}
Let us introduce the $D_0$ covariant derivative:
\[
D_0 = d + J_0 \,.
\]
This is just the standard Levi-Civita metric connection in the tangent space
to $AdS_5\times S^5$. (While the ``full" covariant derivative
$d+J$ can sometimes be identified as the ``long" connection modified
by the Ramond-Ramond five form field strength; it
is roughly speaking 
$d+{1\over 2}\omega^{ab}\Gamma_{ab}+d\hat{x} F^{abcde}\Gamma_{abcde}$).
The tangent space to the space of solutions of (\ref{ZeroCurvature})
is parametrized by $\xi$:
\begin{equation}
        \delta_{\xi}J=d\xi+[J,\xi] \,.
\end{equation}
When $\xi=\xi_3$ we get the equations for $J_1$:
\begin{eqnarray}
&&        D_{0+}J_{1-}+
[J_{3+},J_{2-}]+[J_{2+},J_{3-}]-[N_{0+},J_{1-}]+[J_{1+},N_{0-}]=0 \\[5pt]
&&        D_{0-}J_{1+}+[J_{1-},N_{0+}]-[N_{0-},J_{1+}]=0 \,.
\end{eqnarray}
When $\xi=\xi_1$ we get the equations for $J_3$:
\begin{eqnarray}
&&        D_{0+}J_{3-}-[N_{0+},J_{3-}]+[J_{3+},N_{0-}]=0 \\[5pt]
&&        D_{0-}J_{3+}-[N_{0+},J_{3-}]+[J_{3+},N_{0-}]-
[J_{2+},J_{1-}]-[J_{1+},J_{2-}]=0 \,.
\end{eqnarray}
When $\xi=\xi_2$ we get the equations for $J_2$:
\begin{eqnarray}
&&        D_{0+}J_{2-}+[J_{3+},J_{3-}]-[N_{0+},J_{2-}]+[J_{2+},N_{0-}]=0 \\
&&        D_{0-}J_{2+}-[J_{1+},J_{1-}]-[N_{0+},J_{2-}]+[J_{2+},N_{0-}]=0 \,.
\end{eqnarray}
The pure spinors do not change the condition that
\begin{equation}
\partial_+ J_{0-} - \partial_- J_{0+} + [J_{0+}, J_{0-}] +
[J_{2+}, J_{2-}] + [J_{3+}, J_{1-}] + [J_{1+}, J_{3-}]= 0  \,.
\end{equation}
This is  a ``geometrical condition" on the worldsheet connection.
In deriving the equations of motion for $\lambda$ we have to take into
account that
\begin{equation}
        \mbox{Str}\; \{w_{1+},\lambda_3\} J_{0-} =
        \mbox{Str}\; w_{1+}[\lambda_3, J_{0-}] \,,
\end{equation}
because $w$ and $\lambda$ are both odd elements of the superalgebra.
For odd $a$ and $b$ we have $\mbox{Str}\; ab =-\mbox{Str}\; ba$.
Therefore the equations of motion for $\lambda$ are:
\begin{eqnarray}
&&        D_{0-}\lambda_3 - [N_{0-},\lambda_3] =0 \label{EqOfLambdaOne}\\
&&        D_{0+}\lambda_1 - [N_{0+},\lambda_1]= 0 \label{EqOfLambdaThree}\\
&&        [\lambda_3,N_{0+}]=[\lambda_1,N_{0-}]=0 \label{Kinematical}\,,
\end{eqnarray}
where the last equation is kinematical.
The equations of motion for $w$ are:
\begin{eqnarray}
&&        D_{0-}w_{1+} - [N_{0-},w_{1+}]=0 \\
&&        D_{0+}w_{3-} - [N_{0+},w_{3-}]=0 \,.
\end{eqnarray}
This also implies that
\begin{eqnarray}
&&        D_{0-}N_{0+}-[N_{0-},N_{0+}]=0 \\
&&        D_{0+}N_{0-}-[N_{0+},N_{0-}]=0 \,.
\end{eqnarray}
It is useful to introduce the combined current, which will play the role of the Lax pair,
\begin{eqnarray}
&&        J_+(z)=J_{0+}-N_{0+}+{1\over z} J_{3+}+{1\over z^{2}} J_{2+} +{1\over z^{3}} J_{1+}
+{1\over z^{4}} N_{0+} \label{APlusZ}\\
&&        J_-(z)=J_{0-}-N_{0-}+zJ_{1-}+ z^2 J_{2-}+
z^3 J_{3-}+ z^4 N_{0-} \label{AMinusZ} \,,
\end{eqnarray}
where $z$ is the spectral parameter. The equations of motion can then be written as zero curvature conditions:
\begin{equation}\label{ZeroCurvatureA}
        [\partial_+ + J_+(z)\; , \; \partial_- + J_-(z)]=0 \,.
\end{equation}
An important point to notice is that the Lax pair in the pure spinor formulation is different from the one  in \cite{Bena:2003wd} based on the Metsaev-Tseytlin action \cite{Metsaev:1998it}. This is true even after dropping the ghost terms. The Lax connection in the Metsaev-Tseytlin formulation is
\begin{equation}
J_+^{MT}(z) = J_{0+} + z J_{1+} + {1\over z} J_{3+} + {1\over z^2} J_{2+} \,,
\end{equation}
and the resulting equations are different in the matter sector. 
Nevertheless, the theories are of course classically equivalent, which follows by choosing the specific gauge in the Metsaev-Tseytlin formulation $J_{1+}=0$ and $J_{3-}=0$.

Now let us notice that the equations of motion (\ref{EqOfLambdaOne}), (\ref{EqOfLambdaThree}) and (\ref{Kinematical}) are
equivalent to the statement that the coefficients:
\begin{itemize}
\item         of $z^{-5}$         in $[\partial_++J_+(z),z^{-1}\lambda_3]$,
\item         of $z$        in $[\partial_++J_+(z),z\lambda_1]$,
\item        of $z^{5}$        in $[\partial_-+J_-(z),z\lambda_1]$,
\item        of $z^{-1}$        in $[\partial_-+J_-(z),z^{-1}\lambda_3]$ \,,
\end{itemize}
are all zero.
Therefore the BRST transformation is given by this formula:
\begin{equation}
[ \epsilon Q , J_{\pm}(z) ] = D_{\pm}^{(z)}(\epsilon \lambda(z)) \,,
\end{equation}
where $\lambda(z)={1\over z}\lambda_3 + z\lambda_1$. 
This means that $Q$ acts as an infinitesimal
dressing transformation\footnote{Dressing transformations are
the gauge transformations of the Lax connection preserving the analytical
structure of the connection as a function of the spectral
parameter \cite{ZakharovShabat}.}.

\subsection{The transfer matrix}

The transfer matrix is defined as the path-ordered exponential:
\begin{equation}\label{TheMonodromyMatrix}
\Omega_{\tau_r}^{\tau_l} (z) =  P\exp \left[ 
-\int_{\tau_r}^{\tau_l} \left(J_+(z)d\tau^+ + J_-(z)d\tau^- \right) 
\right] \,.
\end{equation}
The zero-curvature equations (\ref{ZeroCurvatureA}) are equivalent to the 
flatness of the connection $J(z)$; this implies that $\Omega(z)$ does
not depend on the choice of the contour. 
Classically, with the periodic boundary conditions, 
the expansion coefficients of $\mbox{Str}\; \Omega (z)$ in $z$ around 
$0, \infty$  yield an infinite family of local charges \cite{Beisert:2005bm}, 
and the expansion around $z=1$ results in an infinite set of non-local charges.
Classically these charges are all in involution (their Poisson brackets vanish).

Here we will study the logarithmic divergences of $\Omega(z)$ 
by explicitly computing the short-distance expansion of the currents.   

\subsection{``Small case" currents}
\label{sec:SmallCaseCurrents}
The ``small case currents" $j_1 , j_2 , j_3$
are defined as follows:
\begin{equation}
j_a = g^{-1}J_{a}g  \,.
\end{equation}
We also define $j_0$:
\begin{equation}
j_0=g^{-1}N g \,.
\end{equation}
The most important property of these small-case currents is that they
are gauge invariant under (\ref{GaugeTransformationCoset}). 
They do not have a definite grading; it is not true that $j_1$ belongs to 
$\mathfrak{g}_{\bar{1}}$. 
The global conserved charges corresponding to the Killing vectors and spinors of
$AdS_5\times S^5$ are given by linear combinations:
\begin{equation}
q_{global}=\int *(4j_0+ 3j_1 + 2j_2 +j_3) \,.
\end{equation}
L\"uscher used the small case currents to construct the Yangian conserved
charge in the $O(n)$ nonlinear sigma-model \cite{Luscher:1977uq}.
In his approach the Yangian charge was constructed from the ordered
double integral of the small case currents:
\begin{equation}\label{SmallCaseYangian}
q_{2,Yangian} = 
\int_{-\infty}^{+\infty} d\tau_2 \int_{-\infty}^{\tau_2} d\tau_1 \;
j(\tau_2) j(\tau_1) + \int_{-\infty}^{+\infty} d\tau \;k(\tau) \,.
\end{equation}
It should be possible to generalize his arguments and prove the finiteness
of the higher conserved charges for the string in $AdS_5\times S^5$.
The logarithmic divergences were present (and in fact played an important
role) in \cite{Luscher:1977uq, Abdalla:1980jt, Abdalla:1982yd, Evans:2004ur}. They appeared because of the
collisions $\tau_2\to \tau_1$ in (\ref{SmallCaseYangian}), see
Section \ref{sec:DoubleCollisions}. 
But they were all proportional
to algebraic structures like $f_a^{bc}[t_b,t_c]$, which would
vanish for the algebra  $\mathfrak{psu}(2,2|4)$ because the adjoint
Casimir is zero in this case.

In this paper we will take a more pedestrian approach and 
calculate the divergences explicitly using the capital currents
expanded around flat space. Our main reason is the following.
We feel that the transfer matrix itself should play a fundamental role,
rather than the nonlocal conserved charges which are obtained by expanding
around $z=1$. 
In \cite{Luscher:1977uq} only the bilocal charges were explicitly studied
(notice that the higher nonlocal charges could be obtained by calculating
the Poisson brackets of the bilocal charges). It was important that the
only source of logarithmic divergences were double collisions of $j$.
In fact, in the expansion of the transfer matrix, triple collisions
contribute to the divergence already at the one loop level. And
quadruple and higher order collisions would contribute at higher loops.
This does not affect the bilocal charge, but the argument based
on generating the higher order charges by applying the Poisson brackets
to the bilocal charges is rather indirect.

The bilocal charge does not really ``probe" the structure of the Lax
equation (\ref{ZeroCurvatureA}), (\ref{APlusZ}), (\ref{AMinusZ}).
One could imagine that the bilocal conserved charge is given 
by an expression of the form (\ref{SmallCaseYangian}), but this does not yet imply 
that the conserved charge can be obtained from $\Omega(z)$ given
by (\ref{TheMonodromyMatrix}) with $J$ given by (\ref{APlusZ}), (\ref{AMinusZ}).
In the pure spinor formalism the expressions (\ref{APlusZ}), (\ref{AMinusZ})
for the Lax connection appear rather artificial, and it is useful
to verify that this is a sensible construction in the quantum theory,
by explicit calculations.

Note, that in \cite{Berkovits:2004jw} the BRST-invariance of $q_{2,Yangian}$ 
was proven. The argument relied on the vanishing of the ghost-number 
+1 BRST-cohomology class.

\section{Action and OPE in the near-flat space limit}
\label{sec:ActionAndOPE}

The flat-space limit of $AdS_5 \times S^5$ requires taking the large 
radius limit $R \rightarrow \infty$ and rescaling the bosonic and 
fermionic fields with $1\over R$ so that the action becomes
a quadratic expression like $(\partial x)^2 + (\partial\theta)^2$
plus nonlinear terms proportional to powers of $1\over R$.
We will now explain how to do this rescaling. 
 
\subsection{Viel-bein}
A viel-bein is a choice of basis in the tangent space to $AdS_5\times S^5$.
It is equivalent to the choice of the lift
$g\in PSU(2,2|4)$ for each point in
${PSU(2,2|4) \over SO(4,1) \times SO(5)}$.
We will parametrize the points of the space $AdS_5\times S^5$
by  $(x^{\mu},\vartheta_L^{\alpha},\vartheta_R^{\dot{\alpha}})$ as in \cite{Metsaev:1998it}, so that
the viel-bein is:
\begin{equation}\label{VielbeinGaugeFixing}
g(\vartheta, x)    =  
\exp\left(
{1\over R} \vartheta^{\alpha}_L t^3_{\alpha}  
+ {1\over R} \vartheta^{\dot{\alpha}}_R t^1_{\dot{\alpha}}
\right)   
\exp\left(
{1\over R} x^{\mu}t_{\mu}^2 \right) \,.
\end{equation}
Let us introduce the shorthand notations:
\[
 x=x^{\mu}t_{\mu}^2\,, \qquad
\vartheta_L=\vartheta_L^{\alpha}t^3_{\alpha} \qquad \mbox{and} \qquad 
\vartheta_R=\vartheta_R^{\dot{\alpha}}t^1_{\dot{\alpha}} \,.
\]
Notice that $x$,  $\vartheta_L$ and $\vartheta_R$  are all even
elements of the Lie superalgebra $\mathfrak{psu}(2,2|4)$. 
With this notation the viel-bein can be written as follows:
\begin{equation}
g  =  e^{{1\over R}(\vartheta_L+\vartheta_R)} e^{{1\over R}x} \,.
\end{equation}
If $g=\exp \xi^a t_a $ then 
\begin{equation}
\partial g g^{-1} 
= \partial\xi
- {1\over 2} [\partial \xi, \xi]
+ {1\over 6} [[\partial \xi, \xi], \xi]
+\ldots \,.
\end{equation}
This implies the expansion for the currents 
in terms of the flat-space fields $x$ and $\vartheta$:
\begin{eqnarray}
-J_{2+} & = & {1\over R}\partial_+ x
+{1\over 2R^2} [ \vartheta_L , \partial_+ \vartheta_L  ]
+{1\over 2R^2} [ \vartheta_R , \partial_+ \vartheta_R ] +
\nonumber\\
&&
+ {1\over 6R^3} [x , [x, \partial_+ x ]]
+ {1\over 2R^3} [ \vartheta_R , [\vartheta_L , \partial_+ x]  ]
+ {1\over 2R^3} [ \vartheta_L , [\vartheta_R ,  \partial_+ x]]
+ O\left({1\over R^4} \right)
\nonumber \\[5pt]
-J_{3+} &  = & 
 {1\over R} \partial_+\vartheta_L 
+ {1\over R^2} [ \vartheta_R ,\partial_+ x  ]
\nonumber \\ &&
+ {1\over 2R^3} [ \vartheta_L , [x , \partial_+ x] ] 
+ {1\over 6R^3} [ \vartheta_R , [ \vartheta_L , \partial_+ \vartheta_L ] ]
+ {1\over 6R^3} [ \vartheta_L , [ \vartheta_R , \partial_+ \vartheta_L ] ]
\nonumber \\ &&
+ {1\over 6R^3} [ \vartheta_L , [ \vartheta_L , \partial_+ \vartheta_R ] ]
+ {1\over 6R^3} [ \vartheta_R , [ \vartheta_R , \partial_+ \vartheta_R ] ]
+ O\left({1\over R^4} \right)
\nonumber \\[5pt]
-J_{1+}  & = & 
 {1\over R}\partial_+\vartheta_R 
+ {1\over R^2} [ \vartheta_L , \partial_+ x ]
\nonumber \\ &&
+ {1\over 2R^3} [ \vartheta_R ,  [x , \partial_+ x] ] 
+ {1\over 6R^3} [ \vartheta_L , [ \vartheta_L , \partial_+ \vartheta_L ] ]
+ {1\over 6R^3} [ \vartheta_R , [ \vartheta_R , \partial_+ \vartheta_L ] ]
\nonumber \\ &&
+ {1\over 6R^3} [ \vartheta_R , [ \vartheta_L , \partial_+ \vartheta_R ] ]
+ {1\over 6R^3} [ \vartheta_L , [ \vartheta_R , \partial_+ \vartheta_R ] ]
+ O\left({1\over R^4} \right)
\nonumber \\[5pt]
-J_{0+} & = &  
 {1\over 2 R^2} [x , \partial_+ x]
+ {1\over 2 R^2} [ \vartheta_R , \partial_+ \vartheta_L]
+ {1\over 2 R^2} [ \vartheta_L ,  \partial_+ \vartheta_R]
\nonumber \\ &&
+{1\over 2R^3} [ \vartheta_L  , [\vartheta_L , \partial_+ x] ]
+{1\over 2R^3} [ \vartheta_R  , [\vartheta_R , \partial_+ x] ]
+ O\left({1\over R^4} \right)
\,.
\label{JExpansions}
\end{eqnarray}

\subsection{Global symmetries}
\label{sec:GlobalShifts}
Global symmetries act as constant right shifts:
\begin{equation}
g(\vartheta,x)\mapsto g(\vartheta,x)g_0
=h(\vartheta,x; g_0) g(S_{g_0}.(\vartheta,x)) \,,
\end{equation}
where $g_0$ is a constant element of $\mathfrak{psu}(2,2|4)$.
We see that in the gauge (\ref{VielbeinGaugeFixing})
the action of the global symmetry $g_0$ corresponds
to the ``shift" of $\vartheta, x$ 
(which we denoted $S_{g_0}.(\vartheta,x)$)
and a gauge transformation with some parameter $h$ which
is a function of $g_0$ and $x$ (this is sometimes referred to as compensating transformation).
For example, the translation corresponds to
 $g_0=e^{{1\over R}\xi}$ with $\xi\in {\bf g}_2$; 
the corresponding $S_{g_0}$ and  $h$ are:
\begin{eqnarray}
	S_{g_0} x &=& x + \xi + {1\over 3R^2} [x, [x, \xi]] +\ldots
\label{GlobalShiftOfX}\\
         S_{g_0} \vartheta &=& \vartheta + {1\over R^2}[\vartheta , [x, \xi]] + \ldots 
\label{GlobalShiftOfTheta}\\   
 	h(\vartheta,x; e^{\xi}) &=& 
	\exp\left( {1\over 2R^2}[x,\xi] +\ldots\right)
\label{CompensatingGaugeTransformation}
\end{eqnarray}
Therefore in our gauge the global isometries act on the
currents as gauge transformations:
\begin{equation}
S_{g_0}.J = -dh h^{-1} + h J h^{-1}
\end{equation}
where $h=h(\vartheta,x; g_0)$.

\subsection{Action}

The action without the ghost terms is
\begin{equation}
S={1\over \pi}\int d^2v 
\left({1\over 2} C_{\mu\nu}J_{2+}^{\mu}J_{2-}^{\nu} +
{1\over 4} C_{\alpha\dot{\beta}} J_{3+}^{\alpha}J_{1-}^{\dot{\beta}} +
{3\over 4} C_{\dot{\alpha}\beta} J_{1+}^{\dot{\alpha}} J_{3-}^{\beta}
\right) \,.
\end{equation}
Here $C_{\mu\nu}$, $C_{\alpha\dot{\beta}}$ and $C_{\dot{\alpha}\beta}$
are the invariant tensors, see Appendix A.

\noindent
The path integral is the sum over histories of $e^{-S}$.

\noindent
The near-flat space expansion of the currents yields
\begin{eqnarray}\label{ActionNearFlat}
S&=&{1\over\pi}\int d^2v 
\left({1\over 2}  C_{\mu\nu}\partial_+ x^{\mu} \partial_- x^{\nu}
+C_{\dot{\alpha}\beta}
\partial_+\vartheta^{\dot{\alpha}}_R \partial_-\vartheta^{\beta}_L
-\right.\nonumber \\
&& \left.-{1\over 2}{1\over R} f_{\mu\alpha\beta}\partial_+ x^{\mu}
\vartheta_L^{\alpha} \partial_-\vartheta_L^{\beta}
-{1\over 2}{1\over R} f_{\mu\dot{\alpha}\dot{\beta}}\partial_- x^{\mu}
\vartheta_R^{\dot{\alpha}} \partial_+\vartheta_R^{\dot{\beta}}
+\ldots
\right) \,.
\end{eqnarray}
Further we will need to know the order $1/R^2$ terms as well. These are
\begin{equation}
 {1\over \pi} \int d^2 v {1\over R^2} \hbox{Str} \mathcal{L}_2\,,
\end{equation}
where 
\begin{equation}
\begin{aligned}
\mathcal{L}_2 = 
& - {1\over 6}  [ x , \partial_+ x ] [ x ,  \partial_- x ]]  \cr
& - {1\over 4}  [ \vartheta_R , \partial_+ x] [ \vartheta_L , \partial_- x] 
  + {1\over 4}  [ \vartheta_L , \partial_+ x] [ \vartheta_R , \partial_- x] \cr
& - {1\over 8}  [\vartheta_R, \partial_+ \vartheta_L] [x, \partial_- x] 
  - {1\over 8}  [\vartheta_L, \partial_- \vartheta_R] [x, \partial_+ x]\cr
& - {3\over 8}  [\vartheta_L, \partial_+ \vartheta_R] [x, \partial_- x] 
  - {3\over 8}  [\vartheta_R, \partial_- \vartheta_L] [x, \partial_+ x] \cr
&  - {1\over 24} [\vartheta_L , \partial_+ \vartheta_L] 
[\vartheta_L , \partial_- \vartheta_L] 
  - {1\over 24} [\vartheta_R , \partial_- \vartheta_R] 
[\vartheta_R , \partial_+ \vartheta_R] \cr
& + {1\over 24} [\vartheta_L ,\partial_+ \vartheta_L] 
[\vartheta_R , \partial_- \vartheta_R]
  - {1\over 8}  
[\vartheta_L ,\partial_- \vartheta_L] 
[\vartheta_R , \partial_+ \vartheta_R] \cr
& - {1\over 6}  [\vartheta_R ,\partial_+ \vartheta_L] 
[\vartheta_R , \partial_- \vartheta_L]
  - {1\over 6}  [\vartheta_L ,\partial_- \vartheta_R] 
[\vartheta_L , \partial_+ \vartheta_R] \cr
& - {1\over 12} [\vartheta_R ,\partial_+ \vartheta_L] 
[\vartheta_L , \partial_- \vartheta_R]
  - {1\over 4}  [\vartheta_L ,\partial_+ \vartheta_R] 
[\vartheta_R , \partial_- \vartheta_L] \,. 
\end{aligned}
\end{equation}

\subsection{The OPEs of the elementary fields}
From the quadratic part of the action we have
\begin{eqnarray}
\langle x^{\mu}(w,\bar{w})x^{\nu}(0)\rangle & = & -\pi C^{\mu\nu} 
(\partial_w\partial_{\bar{w}})^{-1}
\\
\langle \vartheta_L^{\alpha}(w,\bar{w})\vartheta_R^{\dot{\beta}}(0)\rangle 
& = &
-\pi C^{\alpha\dot{\beta}} 
(\partial_w \partial_{\bar{w}})^{-1}
\label{ElementaryOPE}
\\
\langle \vartheta_R^{\dot{\alpha}}(w,\bar{w})
\vartheta_L^{\beta}(0)\rangle 
& = & 
- \pi C^{\dot{\alpha}\beta}(\partial_w\partial_{\bar{w}})^{-1} \,.
\end{eqnarray}
Notice that 
${\partial\over\partial \bar{w}}{1\over w} = \pi\delta^2(w,\bar{w})$.
Therefore
\begin{equation}
\langle x^{\mu}(w,\bar{w})x^{\nu}(0)\rangle =
- C^{\mu\nu} \log |w|^2\,.
\end{equation}

\subsection{A useful ``symmetry"}
\label{sec:Symmetry}
Notice that all the formulas are valid if $J_+$ is exchanged with $J_-$,
dotted spinor indices exchanged with undotted and $\vartheta_L$ with $\vartheta_R$.
In other words
\begin{equation}
\label{MetaSymmetry}
(J_{0+}^{[\mu\nu]} \leftrightarrow J_{0-}^{[\mu\nu]} \;,\; 
 J_{1+}^{\dot{\alpha}} \leftrightarrow J_{3-}^{\alpha} \;,\;
 J_{2+}^{\mu} \leftrightarrow J_{2-}^{\mu} \;,\;
 J_{3+}^{\alpha} \leftrightarrow J_{1-}^{\dot{\alpha}} )
\end{equation}

\subsection{ OPE of currents}

The near-flat space expansion of the currents and the OPE of $x$ and $\vartheta$ imply the OPE of the currens to order $1/R^3$
\begin{eqnarray}
J_{1+}^{\dot{\alpha}}(w_1)J_{2+}^{\mu}(w_2) 
& = &
 {1\over R^3} { \partial_+ \vartheta_L^{\gamma} \over w_1 - w_2 }
 \fduu{\gamma}{\dot{\alpha}}{\mu} 
+ O\left({1\over R^4}\right)
\\
J_{3+}^{\alpha}(w_3)J_{2+}^{\mu}(w_2)
& = &
{2\over R^3} {  \partial_+\vartheta_R^{\dot{\beta}} \over w_3 - w_2 }
\fduu{\dot{\beta}}{\alpha}{\mu} 
+ {1\over R^3} {\bar{w}_3 -\bar{w}_2 \over (w_3-w_2)^2} 
   \partial_- \vartheta_R^{\dot{\gamma}} f_{\dot\gamma}{}^{\alpha\mu}
+  O\left({1\over R^4}\right)
\label{OPEJ1PlusJ2Plus}
\\
J_{1+}^{\dot{\alpha}}(w_a) J_{1+}^{\dot{\beta}}(w_b) 
& = &
-{1\over R^3} {\partial_+ x^{\mu}\over w_a - w_b }
 \fduu{\mu}{\dot{\alpha}}{\dot{\beta}}
+ O\left({1\over R^4}\right)
\label{OPEJ1PlusJ1Plus}
\\
J_{3+}^{\alpha}(w_a) J_{3+}^{\beta}(w_b) 
& = &
-{2\over R^3} {\partial_+ x^{\mu} \over w_a - w_b }
\fduu{\mu}{\alpha}{\beta}
-{1\over R^3} {\bar{w}_a -\bar{w}_b \over (w_a - w_b)^2 }\partial_- x^{\mu}
\fduu{\mu}{\alpha}{\beta}
+ O\left({1\over R^4}\right)
\label{OPEJ3PlusJ3Plus}
\\
J_{1+}^{\dot{\alpha}}(w_1) J_{3+}^{\alpha}(w_3) 
& = &
-{1\over R^2} {1\over (w_1-w_3)^2}
C^{\dot{\alpha}\alpha} 
+ O\left({1\over R^4}\right)
\\
J_{2+}^{\mu}(w_m) J_{2+}^{\nu}(w_n) 
&=&
-{1\over R^2} {1\over (w_m - w_n)^2 }
C^{\mu\nu} 
+ O\left({1\over R^4}\right)
\\
J_{0+}^{[\mu\nu]}(w_0) J_{1+}^{\dot{\alpha}}(w_1) 
& = &
- {1\over 2 R^3}  \left(
{\vartheta_R^{\dot{\beta}}(w_0) \over (w_0 - w_1)^2}
+
{\partial_+\vartheta_R^{\dot{\beta}}(w_0) \over (w_0 - w_1)}
\right)
\fduu{\!\!\!\!\!\dot{\beta}}{\;\dot{\alpha}}{[\mu\nu]}
+ O\left({1\over R^4}\right)
\label{OPEJ0J1}
\\
J_{0+}^{[\mu\nu]}(w_0) J_{3+}^{\alpha}(w_3) & = &
- {1\over 2 R^3}  \left(
{\vartheta_L^{\beta}(w_0) \over (w_0 - w_3)^2}
+
{\partial_+\vartheta_L^{\beta}(w_0) \over (w_0 - w_3)}
\right)
\fduu{\!\!\!\!\!\beta}{\;\alpha}{[\mu\nu]}
+ O\left({1\over R^4}\right)
\label{OPEJ0J3}
\\
J_{0+}^{[\mu\nu]}(w_0) J_{2+}^{\lambda}(w_2) & = &
- {1\over 2 R^3}  \left(
{x^{\kappa}(w_0) \over (w_0 - w_2)^2}
+
{\partial_+ x^{\kappa}(w_0) \over (w_0 - w_2)}
\right)
\fduu{\!\!\!\!\!\kappa}{\lambda}{[\mu\nu]}
+ O\left({1\over R^4}\right) \,.
\label{OPEJ0J2}
\end{eqnarray}
The OPEs are computed by evaluating all the Feynman diagrams $J_+ J_+$, 
including the non-linear terms
in the action, and thus differ from just evaluating the OPE of $J_+ J_+$ 
using the expansion in (\ref{JExpansions}) and the free field OPEs of $x$ 
and $\vartheta$. This in particular leads to the interesting terms involving 
$\bar{w}$ and the left-moving fields, which are however not unexpected in an 
interacting theory, where the decoupling of left and right-moving modes is 
generically not possible. 
There are analogous OPEs of $J_-J_-$ and $J_+ J_-$.
We will not write the complete table here. 
We will compute the necessary singularities in the following sections
  as we need them.

The OPEs of the currents were computed also in 
 \cite{Puletti:2006vb} using the background field method. 
 Our OPEs are in agreement\footnote{There was a mistake in the original
version of our paper. The mistake was in 
the coefficients of the singularities of $J_1J_2$ and $J_3J_2$, which we are not using in 
our further calculations. Notice the difference of notations:
${\bf g}_{\bar{1}}$ of our paper corresponds to ${\bf g}_{\bar{3}}$ of
\cite{Puletti:2006vb}.} with \cite{Puletti:2006vb}.

\subsection{Field renormalization}
\label{sec:FieldRenormalization}
The worldsheet theory for the pure spinor superstring in
$AdS_5\times S^5$ is believed to be UV finite. But this does not preclude
log divergences, if these
divergences could be absorbed into the field redefinition.
It turns out that 
there is a renormalization of the field $x$, 
 because there are logarithmic divergences 
 of the type $\partial_+x\partial_-x$.
There exist two sorts of contribution to such divergences, the
normal ordering of the quartic vertices and the fish diagram
from the contraction of the two cubic vertices. 
\vspace{10pt}

\begin{centering}
	\hfill
	\epsfxsize=3in
      \epsffile{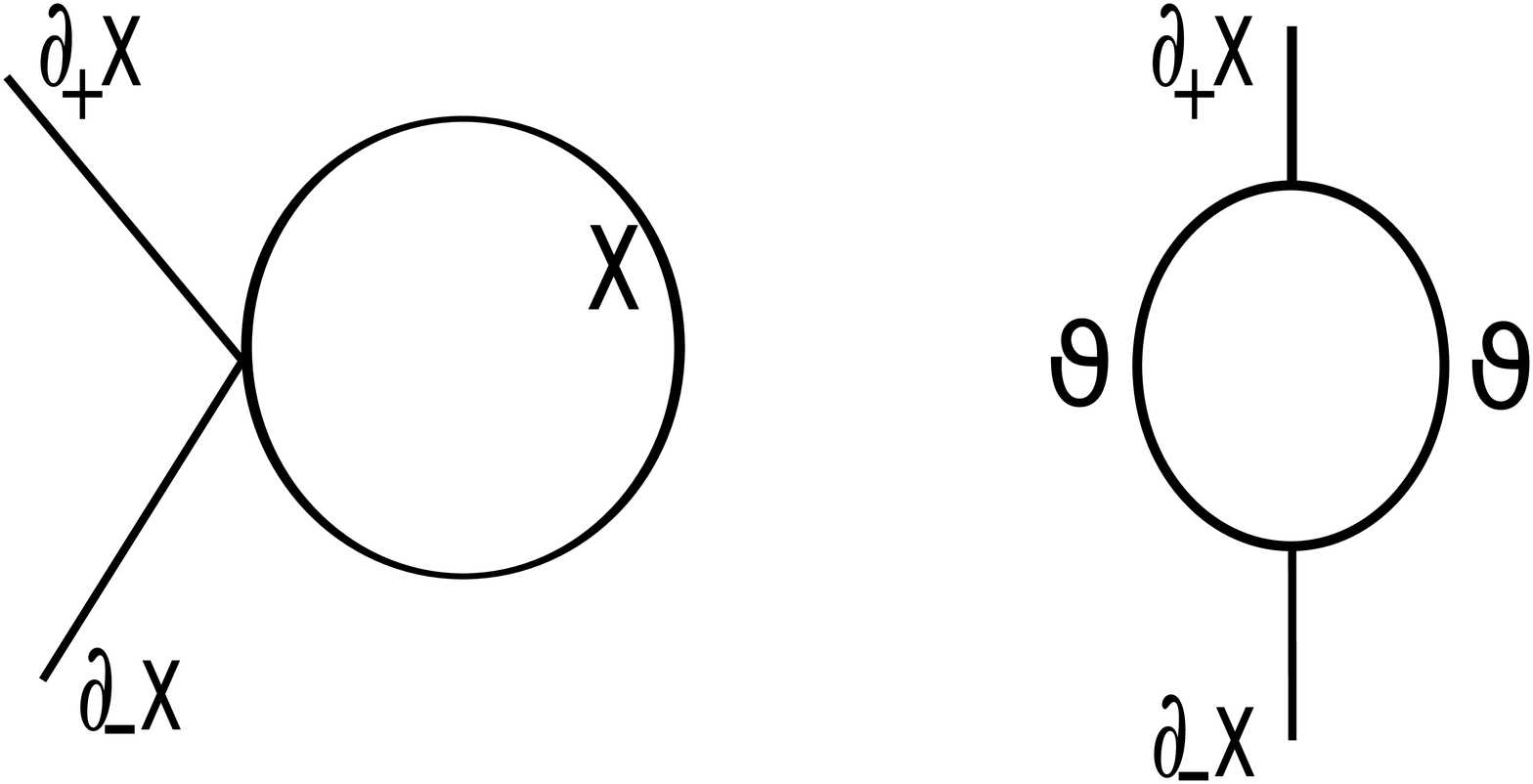}
      \hfill
\end{centering}

\vspace{10pt}

\noindent \underline{\em Quartic vertices.}
 The contribution comes
from the interaction terms in the action of the form 
$x\; x\; \partial x \; \partial x$ and 
$\vartheta\; \vartheta\; \partial x\; \partial x$.
These terms come from the $J_{2+}J_{2-}$, $J_{3+}J_{1-}$
and $J_{1+}J_{3-}$ terms in the action. Our definition
of the group element $e^{R^{-1}(\vartheta_L+\vartheta_R)} e^{R^{-1}x}$
is such that the terms in $J=-dg g^{-1}$ not containing $\partial x$ 
do not contain $x$ at all. Taking this into accont, the terms in 
$J_{2}$ relevant to the wave function renormalization are:
\begin{equation}
J_{2} = -d x - {1\over 2} [\vartheta_L , d\vartheta_L]
-{1\over 2} [\vartheta_R , d\vartheta_R]
-  {1\over 6} [ x, [x , d x] ] -
{1\over 2} [\vartheta_L , [\vartheta_R , d x] ] -
{1\over 2} [\vartheta_R , [\vartheta_L , d x] ] +\ldots
\end{equation}
The quartic terms in the Lagrangian leading to the log divergences
of the type $\partial_+ x \partial_- x$ are:
\begin{eqnarray}
{1\over 6} (\partial_+ x, [ x , [ x, \partial_- x] ] ) 
& + &
{1\over 2} (\partial_+ x, [\vartheta_L , [\vartheta_R , \partial_- x] ] )
+\nonumber
\\
& + &
{1\over 2} (\partial_+ x, [\vartheta_R , [\vartheta_L , \partial_- x] ] )
-\nonumber
\\
& - &
{1\over 4} (\partial_+ x, [\vartheta_R , [\vartheta_L , \partial_- x] ] )
-\nonumber
\\
& - &
{3\over 4} (\partial_+ x, [\vartheta_L , [\vartheta_R , \partial_- x] ] ) \,.
\nonumber
\end{eqnarray}
The log divergences in the terms with fermions cancel, and the
$x\;x\;\partial x\;\partial x $ vertex leads to the log divergence 
\[-{1\over 6}\log\epsilon^2 \;
 (\partial_+ x, [t^{\mu} , [t^{\mu} , \partial_- x] ] )\,.
\] 
which contributes to the renormalization of the field $x$.

\vspace{10pt}
\noindent
\underline{\em Fish diagram.}
The cubic vertices are
$-{1\over 2} ([\partial_+ x ,  \vartheta_L] , \partial_-\vartheta_L )$,
and 
$-{1\over 2}([\partial_- x ,  \vartheta_R] , \partial_+\vartheta_R )$.
The log divergence in the fish is effectively the same as the log 
divergence of the expression
$-{1\over 2}
([\partial_+ x ,  \vartheta_L] , [\partial_- x, \vartheta_R] )$.

\vspace{10pt}
\noindent
Therefore the \underline{total log divergence} 
from the quartic vertices and from
the fish is:
\begin{equation}
-{1\over 6} {1\over R^2}\log\epsilon^2 \;
 (\partial_+ x, [t^2_{\mu} , [t^2_{\mu} , \partial_- x] ] )
-{1\over 2} {1\over R^2}\log\epsilon^2\; C^{\alpha\dot{\beta}} 
(\partial_+ x , \{ t^3_{\alpha} , [ t^1_{\dot{\beta}}, \partial_- x] \}) \,.
\end{equation}
This means that we should replace $x$ with the renormalized $x$, which
is:
\begin{equation}\label{RenormalizationOfX}
x=x^{ren}+{1\over 6} {1\over R^2}\log\epsilon^2\;
[t^2_{\mu} , [t^2_{\mu} , x^{ren}] ]
+{1\over 2} {1\over R^2}\log\epsilon^2\; C^{\alpha\dot{\beta}}
\{ t^3_{\alpha} , [ t^1_{\dot{\beta}}, x^{ren}] \} 
=x^{ren} - {1\over 3 R^2} \log |\epsilon|^2 C_{\bar{2}}.x^{ren}
\,.
\end{equation} 
In other words, the renormalization of $dx$ is such that
this expression:
\begin{equation}
dx + {1\over 6}  [ x , [ x, d x] ]  
+ {1\over 2}[\vartheta_L , [\vartheta_R , d x ]] \,.
\end{equation}
remains finite.
One can see that this is the same as saying that $J_{2}$ is finite.
(We should stress that our analysis is only valid to the order $R^{-3}$.)

Eq. (\ref{RenormalizationOfX}) can be checked against the formula
(\ref{GlobalShiftOfX}) for the global shift. The expression for
$S_{g_0}x^{ren}$ as a function of $x^{ren}$ is the same as the 
expression for $S_{g_0}x$ as a function of $x$: 
\begin{equation}\label{GlobalShiftNonRen}
S_{g_0} x^{ren} = x^{ren}+\xi + {1\over 3R^2}:[x^{ren},[x^{ren},\xi]]:+\ldots
\end{equation}
if we take into account that 
$[x,[x,\xi]]=:[x,[x,\xi]]:-\log|\epsilon|^2 C_{\bar{2}}.\xi+\ldots$.
This agrees with the non-renormalization of $R$.

\section{Renormalization of the Wilson line type of operators}
\label{sec:RenormalizationWilsonLine}

\subsection{Logarithmic divergences}

We consider a nonlocal operator of the form 
\begin{equation}
\Omega[\Gamma]=P\exp \left( -\int_{\Gamma} J \right)\,,
\end{equation}
where $\Gamma$ is a contour
and $J$ has a regular expansion in powers of $R^{-1}$,
starting with the leading term of the order $R^{-1}$.
We assume that $J$ is a 1-form and an element of the Lie superalgebra
(in our case $\psu(2,2|4)$).
We expand the path ordered exponential in powers of $R^{-1}$
and get an infinite series of terms of the type
\begin{equation}\label{ToBeRenormalized}
\int_{\tau_1<\tau_2<\ldots<\tau_n} J(\tau_1) \cdots J(\tau_n) \,.
\end{equation}
When we compute the expectation value of this operator 
we typically encounter linear and logarithmic divergences.
Linear divergences depend on the regularization scheme.
But logarithmic divergences should not depend on the regularization
scheme. 
 
A Wilson loop type of operator can not be conformally invariant,
and therefore cannot be independent of the choice of 
the contour,
 if logarithmic divergences are present. Indeed, let us consider
two contours $\Gamma$ and $\Gamma'$ which are related by 
a dilatation.

\vspace{10pt}

\begin{centering}
	\hfill
	\epsfxsize=1.3in
      \epsffile{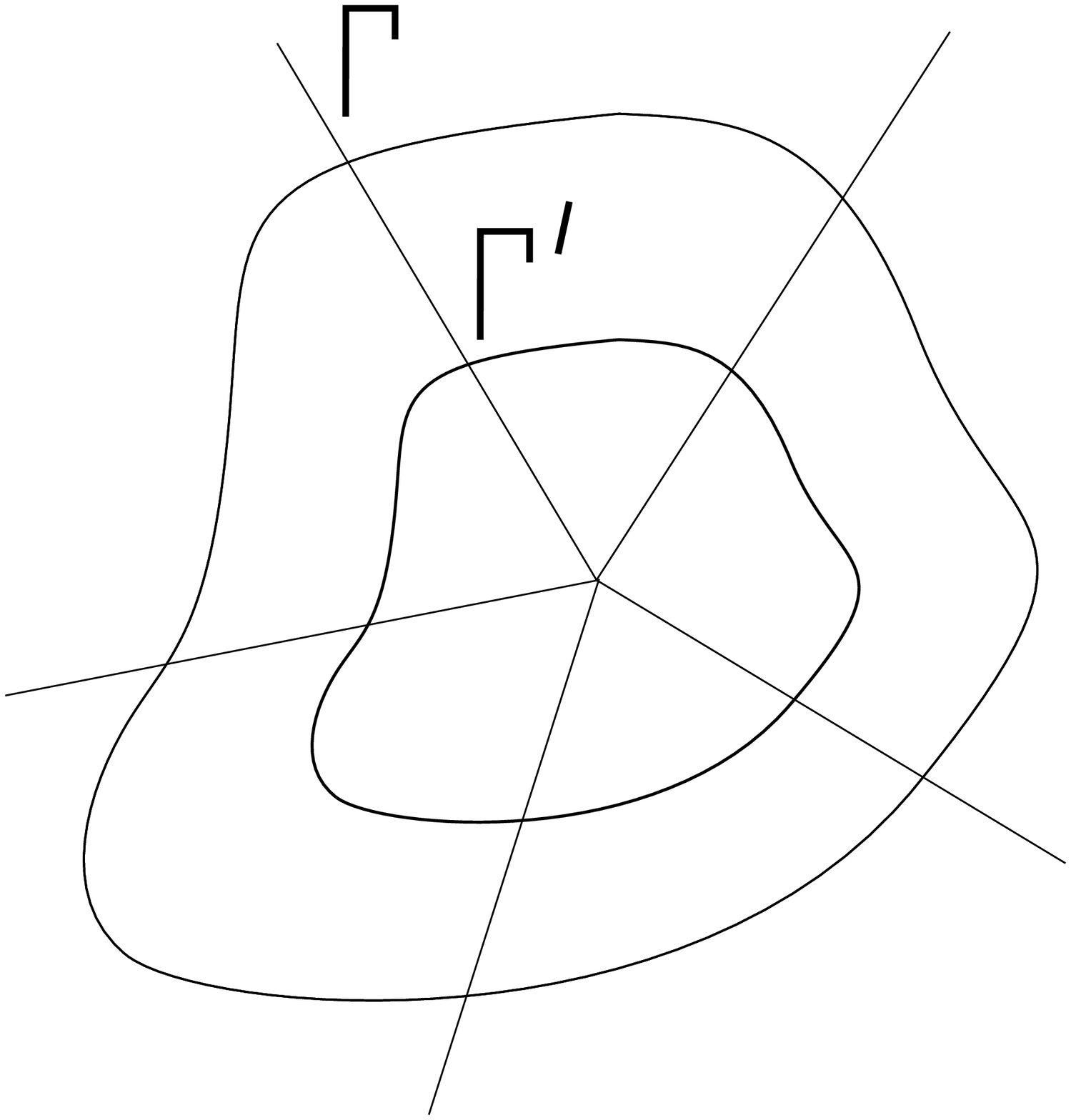}
      \hfill
\end{centering}
\vspace{10pt}

\noindent
This means that $\Gamma'=\lambda \Gamma$ pointwise, where $\lambda$ is a
real number (the dilatation parameter). Both $\Omega[\Gamma]$ and 
$\Omega[\Gamma']$ have UV divergences, therefore we should regularize them.
Let us use the contour split prescription,
as in Appendix B. 
We should use the same regularization for both contours,
with the same parameter $\epsilon$. The renormalized 
Wilson line $\Omega^{ren}$
is equal to the regularized $\Omega_{\epsilon}$ plus counterterms:
\begin{equation}\label{Wren}
\Omega^{ren}[\Gamma] = \lim_{\epsilon\to 0} 
(\Omega_{\epsilon}[\Gamma]+C_{\epsilon}[\Gamma]) \,,
\end{equation}
where $C_{\epsilon}[\Gamma]$ includes all the counterterms.
If it is true that the Wilson line
does not depend on the choice of the contour, then we should have
\begin{equation}\label{WrenDilatationInvariance}
\Omega^{ren}[\Gamma'] = \Omega^{ren}[\Gamma] \,.
\end{equation} 
But let us apply a
dilatation to Eq. (\ref{Wren}). Conformal invariance implies:
\begin{equation}\label{OmegaConformalInvariance}
\Omega_{\epsilon}[\Gamma]=\Omega_{\lambda\epsilon}[\Gamma'] \,.
\end{equation}
By definition $\Omega^{ren}[\Gamma]$ does not depend on $\epsilon$.
If it were true that 
$C_{\lambda\epsilon}[\Gamma']=C_{\epsilon}[\Gamma]$
then (\ref{OmegaConformalInvariance}) 
would imply (\ref{WrenDilatationInvariance}).
But in fact, if there are logarithmic divergences, then it is not 
true that 
$C_{\lambda\epsilon}[\Gamma']=C_{\epsilon}[\Gamma]$.
Indeed, the logarithmic divergences are of the form
\begin{equation}\label{LogNotInvariant}
 \int d\tau^+ \partial_+x \; f(x) \log\epsilon \,,
\end{equation}
and this expression is not invariant under conformal
transformations (because $\epsilon$ has weight 1). 
Notice that the linear divergences, which are
of the form
\begin{equation}
\int {d\tau^+\over \epsilon^+} f(x) \,,
\end{equation}
are invariant under conformal transformations,
but the logarithmic divergences (\ref{LogNotInvariant})
are not.
Therefore the independence of the Wilson line on the
contour should imply the absence of logarithmic
divergences. It is not clear to us whether the converse is
also true.

There are several possible sources of logarithmic divergences.
The currents $A$ are typically composite objects and could therefore
get ``internal" logarithmic divergences\footnote{For example, in the
free theory the ``composite" vertex operator $e^{ik\phi}$ gets
the anomalous dimension $\simeq k^2$ due to its ``internal" divergence}. 
Even if $J$ were elementary fields, in an interacting
theory they could get logarithmic
divergences because of field renormalization. 
Also, logarithmic divergences arise when two or more
points on the integration contour collide, for example 
$\tau_1\to \tau_2$ or $\tau_1 \to \tau_2 \to \tau_3$.

The divergences which we analyze in this paper are the ``internal"
or ``field renormalization" divergences, and the divergences
due to double or triple collisions on the contour. 
Let us first discuss the logarithmic 
divergences due to the double collisions.

\subsection{Double collisions}
\label{sec:DoubleCollisions}
\subsubsection{Double collisions with the second order pole a c-number}
We will now discuss the log divergences in the double collisions
of the type $j_+\llra j_+$. There are also logarithmic
divergences in the collisions $j_+\llra j_-$ and $j_-\llra j_-$.
Suppose that we have the currents $j_+^a(w)$ with the OPE
\begin{equation}
 j_+^a(w) j_+^b(0) =C^{ab}{1\over w^2}+ 
F^{ab}_c k_+^c(0) {1\over w} +
\widetilde{F}^{ab}_c {\bar{w}\over w^2} k_-^c(0) 
+ \ldots 
\end{equation}
where $k_+$ are some other currents,
 $F^{ab}_c=-F^{ba}_c$ are the coefficients of the 
singular term in the OPE and dots denote finite terms.
We will assume for a moment that $C^{ab}$ are c-numbers,
then the second order pole $1\over w^2$ does not contribute
to the logarithmic divergence. Only the simple pole contributes.
Let us consider the path ordered contour integral:
\begin{equation}
\int_{-\infty}^{\infty} dw_1\; j_+^a(w_1) t_a \int_{-\infty}^{w_1}
dw_2\; j_+^b(w_2) t_b \,,
\end{equation}
where $[t_a,t_b]=f_{ab}^c t_c$.
When $w_1\to w_2$ we get the logarithmic divergence of the form:
\begin{equation}
-{1\over 2}\ln\epsilon \int_{-\infty}^{\infty} 
(dw\; j_+^a F^{bc}_a + d\overline{w}\; j_-^a \widetilde{F}^{bc}_a) 
f_{bc}^e t_e \,.
\end{equation}

\subsubsection{Double collisions with field dependent second
order pole}
\label{sec:DoubleCollisionsFieldDependent}
The coefficient of the second order
pole may depend on $w,\bar{w}$, as in (\ref{OPEJ0J1}) --- (\ref{OPEJ0J2}). 
In this case there are two ways of presenting the singularity:
\begin{eqnarray}
j_+^a(w_1) j_+^b(w_2) & = & {C^{ab}(w_1,\bar{w}_1) \over (w_1-w_2)^2}+ 
F^{ab}_c k_+^c{1\over w_1-w_2} +
\widetilde{F}^{ab}_c k_-^c{\bar{w}_1-\bar{w}_2\over (w_1-w_2)^2} +
\ldots = \label{OPEFirstWayOfWriting} \\
& = & {C^{ab}(w_2,\bar{w}_2) \over (w_1-w_2)^2}+
(F^{ab}_c k_+^c + \partial C^{ab}) {1\over w_1-w_2}+
\label{OPESecondWayOfWriting}
\\
&& \spacemaker{{C^{ab}(w_2,\bar{w}_2)\over (w_1-w_2)^2}\;\;\;\;\;}
+(\widetilde{F}^{ab}_c k_-^c + \overline{\partial} C^{ab}) 
{\bar{w}_1 - \bar{w}_2\over (w_1-w_2)^2}
+\ldots
\nonumber
\end{eqnarray}
The path ordered double integral $\int_{w_1>w_2} j_+(w_1)dw_1 \; j_+(w_2)dw_2$
is logarithmically divergent because of the collision $w_1\to w_2$. 
The logarithmic divergence depends on the order of evaluation of the integrals.
Suppose that we first integrate over $w_2$, and then over $w_1$. Then, using
the first formula (\ref{OPEFirstWayOfWriting}) we get the log divergence:
\begin{equation}\label{FirstLogDiv}
-{1\over 2}\int  \log\epsilon \;[t_a,t_b] 
(F^{ab}_c k_+^c(w,\bar{w})dw + \widetilde{F}^{ab}_c k_-^c(w,\bar{w})d\bar{w}) 
\end{equation}
On the other hand, if we first integrate over $w_1$ and then over $w_2$, then 
using the second formula (\ref{OPESecondWayOfWriting}) we get:
\begin{equation}\label{SecondLogDiv}
-{1\over 2}\int  \log\epsilon \; [t_a,t_b]
(F^{ab}_c k_+^c(w,\bar{w})dw + \widetilde{F}^{ab}_c k_-^c(w,\bar{w})d\bar{w}
+dC^{ab}) 
\end{equation}
Notice that the difference between (\ref{FirstLogDiv}) and (\ref{SecondLogDiv})
 is a total derivative. 
It can be integrated and contributes only through the contact terms. To understand
these contact terms, consider for example the ordered integral of three currents:
\begin{equation}
\int_{w_1>w_2>w_3} dw_1 j_+^a(w_1)t_a \;\; dw_2 j_+^b(w_2)t_b \;\; dw_3 j_+^c(w_3)t_c
\end{equation}
Consider the log divergence coming from the collision of $w_1$ with $w_2$.
The difference between (\ref{FirstLogDiv}) and (\ref{SecondLogDiv})
is equal to $[t_a,t_b]dC^{ab}$. This is a total derivative, it almost integrates
to zero except for contact term arising from the condition $w_2>w_3$; a similar
contact term arises when we first collide $w_2$ and $w_3$, and together they give:
\begin{equation}\label{DifferenceOfTwoOrdersOfIntegration}
\int dw_3 [ [t_a,t_b] C^{ab}(w_3) , j_+^c(w_3)t_c]
\end{equation}
But in fact the double pole $1/w^2$ in the product $j^a j^b$ leads to
an additional log divergence from  the {\em triple collision}
$j^a j^b j^c$. This will be explained in the next subsection. If the coefficient
of the double pole is not a constant, then the log divergence of the triple collision
will also depend on the order of integrations. And the difference in the log divergence
of the triple collision calculated with two different orders of integration will
precisely cancel (\ref{DifferenceOfTwoOrdersOfIntegration}).

\subsection{Triple collisions}
\label{sec:TripleCollisions}
\subsubsection{Triple collisions when the second order pole is a c-number}
Triple collisions lead to logarithmic divergences in the case
when there is a second order pole in the OPE $j_+ j_+$.
Let us first assume that the second order pole comes with the
c-number coefficient:
\begin{equation}
j_+^a(w) j_+^b(0) = - {1\over R^2} {1\over w^2} C^{ab} +\ldots   
\end{equation}
The simple pole does not contribute to the logarithmic divergence 
in the triple collisions, and therefore in our discussion of the 
triple collisions we can assume that the singularity
in the product of two currents is just the second order pole, 
as if $j_+^a=\partial_+\phi^a$ with free fields $\phi^a$:
\[
j_+^a(w) j_+^b(0) = - {1\over R^2} {1\over w^2} C^{ab} + :j_+^a(w) j_+^b(0): 
\]
Just to understand how the log divergence appears in triple collisions, 
let us first consider the situation where $n$ constant matrices
$X_1,\ldots, X_n$ are inserted at the positions $a_1,\ldots, a_n$
on the contour:
\begin{equation}
W = P\left[X_1(a_1) \cdots X_n (a_n) \, {\rm exp} 
\left( \int j_+ d\tau^+ \right)  \right] \,.
\end{equation}
We assume that $X_j$ are constant c-number matrices, and the
notation $X_j(a_j)$ just means that $X_j$ is inserted at the point $a_j$
on the contour. This is only needed to specify the right order of
 multiplication of matrices.
Wick's theorem implies
\[
W  =  P\left[X_1(a_1) \cdots X_n (a_n) \, \exp 
      \left(- {1\over 2 R^2} \int d\tau_1^+ \int d\tau_2^+   
 { C^{ab} t^a(\tau_1^+) t^b(\tau_2^+) \over (\tau_1^+ -\tau_2^+)^2  } \right)  
        :\exp \left( \int j_+ d\tau^+\right):  \right] \,.
\] 
Again, $t^a$ are constant c-number matrices (the generators of the
algebra), but we use the notation $t^a(\tau)$ to indicate that $t^a$ is 
inserted at the point $\tau$ on the contour. This is important because
of the path ordering of the product of the matrices. 
Let us consider the effect of just one contraction:
\[
W = P\left[X_1(a_1) \cdots X_n (a_n) 
      \left(- {1\over 2 R^2} \int d\tau_1^+ \int d\tau_2^+   
{C^{ab} t^a(\tau_1^+) t^b(\tau_2^+) \over (\tau_1^+ -\tau_2^+)^2  } \right)  
:\exp \left( \int j_+ d\tau^+\right): \right] \,.
\]
 Let us first illustrate the main point by focusing on four insertions. 
There are three types of terms. 
The first type has both $t^a$ in the same interval between two consecutive
$X$-insertions:
\begin{equation}
 X_{i} \llra t^a \llra t^b \llra  X_{i+1} : 
 \quad   \int_{a_i+ 2 \epsilon}^{a_{i+1} -\epsilon} d\tau_2 
  \int_{a_i + \epsilon}^{\tau_2 - \epsilon} d\tau_1
  {1\over (\tau_1 - \tau_2)^2} = + \log \epsilon \,,
\end{equation}
where we again dropped all subleading terms in $\epsilon$. 
The second possibility is that they are between consecutive insertions
\begin{equation}
X_{i} \llra t^a  \llra X_{i+1} \llra t^b  \llra X_{i+2}: 
 \quad   \int_{a_i+ \epsilon}^{a_{i+1} -\epsilon} d\tau_1 \int_{a_{i+1} + \epsilon}^{a_{i+2}- \epsilon} d\tau_2 {1\over (\tau_1 - \tau_2)^2} = - \log \epsilon \,,
\end{equation}
Finally, all further separated insertions do not contribute 
to the $\log(\epsilon)$ terms:
\begin{equation}
X_{i} \leftrightarrow t^a  \leftrightarrow X_{i+1}\cdots  X_{i+k} \leftrightarrow t^b 
\leftrightarrow  X_{i+k+1}:  \quad   \int_{a_i+ \epsilon}^{a_{i+1} -\epsilon} d\tau_1 
\int_{a_{i+k} + \epsilon}^{a_{i+k+1}- \epsilon} d\tau_2 {1\over (\tau_1 - \tau_2)^2} = 
\mbox{finite} \,.
\end{equation}
With this rule, we can compute the contribution from $J_{2+} J_{2+}$
\begin{equation}\label{FourInsert}
\begin{aligned}
& P\left(\prod_{j=1}^n X_j(a_j)
 \int \int { C^{ab} t^a(\tau_1) t^b(\tau_2) \over (\tau_1^+ - \tau_2^+)^2} \right) \cr
& = \log(\epsilon)  C^{ab} \sum_j
  \cdots X_{j-2}X_{j-1} (t^a t^b X_j - t^a X_j t^b)  X_{j+1} X_{j+3} \cdots
 \cr
& 
=  \sum_j 
\cdots X_{j-2}X_{j-1} \left({1\over 2}\log\epsilon \; C^{ab}[t^a, [t^b, X_j]]\right)
 X_{j+1} X_{j+2} \cdots \,.
\end{aligned}
\end{equation}
Straightforward generalization of this argument yields 
that to lowest order in $1\over R^2$ the logarithmic divergence due
to triple collisions from bosonic currents, in particular $j_+ = J_{2+}$, is given by
\begin{eqnarray}
&& 
P\left[X_1(a_1) \cdots X_n (a_n) \; \exp 
\left( \int j_+ d\tau^+ \right)  \right] 
\nonumber
\\
&&	
\qquad 
\longrightarrow -{1\over 2}{1\over R^2}\log\epsilon \left(
\sum_{k=1}^n 
P\left[ X_1\cdots  C^{ab}  [t^a , [ t^b , X_k ] ] \cdots X_n 
\exp \left( \int j_+ d\tau^+ \right) \right] +\right. \nonumber
\\
&&	
\hspace{80pt}+\left. P\left[ X_1 \cdots X_n 
\int d\tau^+  C^{ab}  [t^a , [t^b , j_+ ] ]
\exp \left( \int j_+ d\tau^+ \right) \right]
\right) \,.
\label{TripleBose}
\end{eqnarray}
Similarly the effect of fermionic currents $j_+$ can be analyzed. 
The OPE has leading order, which is again a c-number
\begin{equation}\label{jjOPE}
j_+^\alpha(w) j_+^{\dot{\beta}}(0) = - {1\over R^2} {1\over w^2} C^{\alpha\dot{\beta}} +\ldots   \,,
\end{equation}
Assuming that the c-number insertions $X_{i}(a_i)$ are bosonic the logarithmic divergence is
\begin{eqnarray}
&& 
P\left[X_1(a_1) \cdots X_n (a_n) \; \exp 
\left( \int j_+ d\tau^+ \right)  \right] 
\nonumber
\\
&&	
\qquad \longrightarrow -{1\over 2}{1\over R^2}\log\epsilon \left(
\sum_{k=1}^n 
P\left[ X_1\cdots    C^{\alpha \dot{\beta}} \{t^{\alpha} , [ t^{\dot{\beta}} , X_k ] \} \cdots X_n 
\exp \left( \int j_+ d\tau^+ \right) \right] +\right.\nonumber\\
&&	
\hspace{80pt}+\left. P\left[ X_1 \cdots X_n 
\int d\tau^+   C^{\alpha \dot{\beta}} [ t^\alpha , \{t^{\dot{\beta}} , j_+ ] \}
\exp \left( \int j_+ d\tau^+ \right) \right]
\right) \,.
\label{TripleFermi}
\end{eqnarray}
Similar expressions hold with suitably altered
 commutators/anti-commutators for fermionic insertions $X_i(a_i)$. 

There are three types of sources for triple-collisions:
 $J_{2+}^\mu J_{2+}^\nu$, $J_{3+}^{\alpha}J_{1+}^{\dot \beta}$ and 
$J_{1+}^{\dot \beta} J_{3+}^{\alpha}$. 
For $X \in {\mathfrak{g}}_{\bar 1} \oplus {\mathfrak{g}}_{\bar 3}$ the Lemma (\ref{NoSpinorDiv}) 
implies vanishing of the logarithms. So the only contributions arise 
from bosonic insertions $X$. 
In this section we will be interested in the transfer matrix itself and thus
 not discuss any insertions. Then the non-trivial contributions are the
 second lines of (\ref{TripleBose}) and (\ref{TripleFermi}).

\subsubsection{Triple collisions with field dependent second order pole}
\label{sec:TripleCollisionsFieldDependent}
So far we considered the case when the most singular
double pole term in the OPE is field-independent, just a c-number.
But in fact we will encounter more general situations, for example in the 
computation of the divergence proportional to $[x , d x]$ in the triple 
collision 
$J_{2+}(0) \leftrightarrow J_{2+}(w_2) \leftrightarrow J_{0+}(w_0)$. 
After evaluating the OPE $J_{2+}(w_2) J_{0+}(w_0)$ one needs to integrate 
$x(w_0)/(w_2-w_0)^2$ with respect to $w_2$ and $w_0$. 
This is the type of triple collision which we will now discuss. 
Just to have a simple example, consider again the system of ``free currents":
\[
j_+^a(w) j_+^b(0) = - {1\over R^2} {1\over w^2} C^{ab} + :j_+^a(w) j_+^b(0): 
\]
Imagine we have some function $\Phi^{ab}(w,\bar{w})$ and consider the path 
ordered double integral:
\[
\int_{w_1>w_2} (t_a\Phi^{ac}(w_1,\bar{w}_1)j_+^c(w_1,\bar{w}_1) dw_1) \;\;\;
(t_b j_+^b(w_2,\bar{w}_2) dw_2)
\]
It is best to think of $\Phi^{ab}(w,\bar{w})$ as a c-number valued function
on the worldsheet. Notice that in our application $J_{0+}={1\over 2}[\partial_+x,x]$
and $\Phi$ is actually a field: $\Phi^{ab}=(\mbox{ad}(x))^{ab}$.  
But for understanding what is going on, it is enough to consider the example
where $\Phi$ is a c-number valued function of $w$ and $\bar{w}$.

The OPE between $\Phi j_+$ and $j_+$ contains a field-dependent second 
order pole  
\begin{equation}\label{FieldDependentSecondOrderPole}
\Phi^{ac}(w_1) j^c_+(w_1) \;\; j^b_+ (w_2) = 
-{1\over R^2 } {\Phi^{ac}(w_1) C^{cb} \over (w_1-w_2)^2} + 
    \Phi^{ac}(w_1) :j^c_+(w) j^b_+ (w_2):
\end{equation}
As we explained in Section \ref{sec:DoubleCollisionsFieldDependent}, the log
divergence from the double collisions depends in this situation on the order
of taking integrals. The log divergence from the triple collisions also depends
on the order of integrations, so that the {\em sum} of the divergences in double
and triple collisions is independent of the order of integrations. 

{\em Let us agree that having the singularity of the form 
(\ref{FieldDependentSecondOrderPole}), we first integrate over $w_2$ and then
over $w_1$.} Then there is no log divergence from the double collision,
but the triple collisions do contribute to the log divergence.  
Let us consider the log divergences coming from the following three integrals 
in triple collisions.

\noindent First integral:
\begin{equation}\label{TripleCollisionFirstIntegral}
X_{i+1} \longleftrightarrow  t^b \Phi^{ba}\longleftrightarrow  t^a 
\longleftrightarrow  X_i:
 \quad   \int_{a_i+ 2 \epsilon}^{a_{i+1} -\epsilon} d\tau_2 
  \int_{a_i + \epsilon}^{\tau_2 - \epsilon} d\tau_1
  { \Phi^{ba}(\tau_2) \over (\tau_1 - \tau_2)^2} = 
 \log \epsilon\;  \Phi^{ba}(a_i)\,.
\end{equation}
Second integral:
\begin{equation}\label{TripleCollisionSecondIntegral}
X_{i+1} \longleftrightarrow t^a \longleftrightarrow  t^b \Phi^{ba} 
\longleftrightarrow  X_i:
 \quad   \int_{a_i+ 2 \epsilon}^{a_{i+1} -\epsilon} d\tau_2 
  \int_{a_i + \epsilon}^{\tau_2 - \epsilon} d\tau_1
  { \Phi^{ba}(\tau_1) \over (\tau_1 - \tau_2)^2} = 
 \log \epsilon\;  \Phi^{ba}(a_{i+1})\,.
\end{equation}
Notice the difference of (\ref{TripleCollisionFirstIntegral})
and (\ref{TripleCollisionSecondIntegral}):
$\log\epsilon\; \Phi(a_i)$ vs. $\log\epsilon\;\Phi(a_{i+1})$.
In (\ref{TripleCollisionFirstIntegral}) we first integrated over $\tau_1$ and
then over $\tau_2$, and in (\ref{TripleCollisionSecondIntegral}) first over 
$\tau_2$ and then over $\tau_1$. 

\noindent Third integral:
\begin{equation}
X_i \leftrightarrow  t^b \Phi^{ba}\leftrightarrow  X_{i+1}  
\leftrightarrow t^a \leftrightarrow X_{i+2}:
\quad \int_{a_{i+1}+\epsilon}^{a_{i}-\epsilon} d\tau_2 
\int_{a_{i+2}+\epsilon}^{a_{i+1}-\epsilon} d\tau_1 
{ \Phi^{ba}(\tau_1) \over (\tau_1 - \tau_2)^2} = -\log\epsilon \;
\Phi^{ba}(a_{i+1}) \,.
\end{equation}
There is also a contribution from the collision
$X_i \longleftrightarrow t^a \longleftrightarrow  X_{i+1}  
\longleftrightarrow   t^b \Phi^{ba}\longleftrightarrow X_{i+2}$
which is also proportional to $-\log\epsilon \;\Phi^{ba}(a_{i+1})$.
The field-dependent triple-collision is therefore
\begin{eqnarray} 
&&P\left [X_1(a_1) \cdots X_n(a_n) 
   \exp \left( \int  j_+^a t^a  d\tau^+  +  \int  \Phi^{ab}  j_+^b  t^a d\tau^+   
\right)  \right] \nonumber\\
&& 
\longrightarrow -{1\over R^2}\log\epsilon \left(
\sum_{k=1}^n 
P\left[ X_1\cdots   [ \Phi^{ab}t^a , [ t^b , X_k ] ] \cdots X_n 
\exp \left( \int j_+ d\tau^+  +  \int  \Phi^{ab}  j_+^b  t_a d\tau^+\right) \right] 
+\right. 
\nonumber
\\
&&	
\hspace{80pt}+\left. P\left[ X_1 \cdots X_n 
\int d\tau^+    [\Phi^{ab}t^a , [t^b , j_+ ] ]
\exp \left( \int j_+ d\tau^+ \right) \right]
\right) \,.
\label{FieldInsertWrapping}
\end{eqnarray}

\subsection{Divergences due to the interaction terms in the action}
We will explain this type of divergences using a simplified model.
Consider a couple of scalar fields $\phi_1$ and $\phi_2$
with the action
\begin{equation}
{1\over \pi}
\int d\tau^+ d\tau^- 
\left(\partial_+ \phi_1 \partial_-\phi_2
+{1\over R} n_+\phi_1\partial_-\phi_2\right)\,,
\end{equation}
where $n_+$ is some function of $\tau^+,\tau^-$; we could treat 
it as a classical source. 

\subsubsection{Effect of the interaction on the double collisions}
\label{sec:EffectOnDoubleCollisions}
\noindent \underline{\bf 1. $(++)\to(+)$} Consider the ordered integral:
\begin{equation}
\int\int_{w_1>w_2}
d w_{1} \; \partial_{w_1}\phi_1 \;\;\; d w_{2} \; \partial_{w_2}\phi_2\,.
\end{equation}
This integral is logarithmically divergent because of the
interaction term in the action. 

\vspace{10pt}
\begin{centering}
	\hfill
	\epsfxsize=2in
      \epsffile{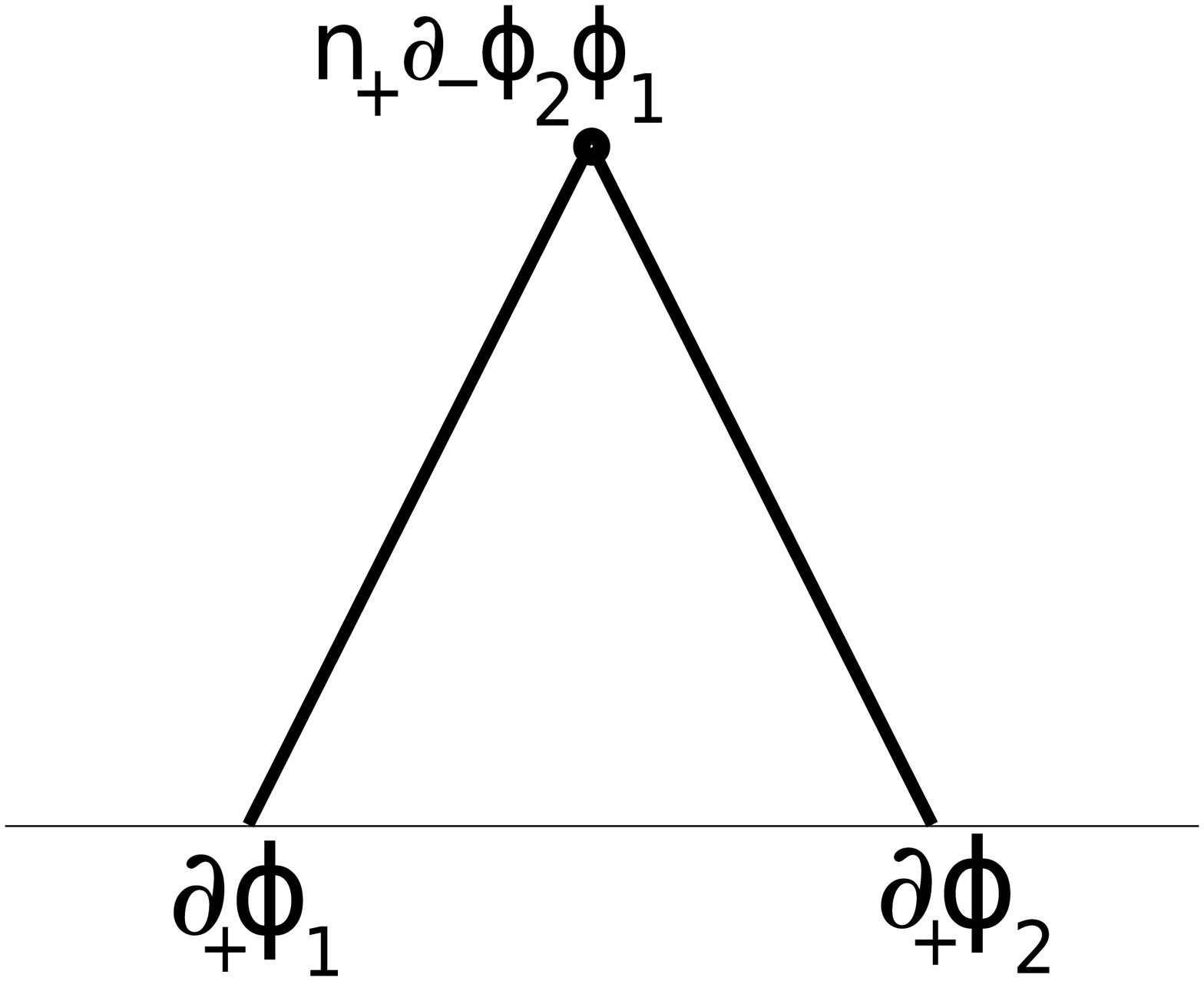}
      \hfill
\end{centering}

\vspace{10pt}
\noindent
Notice that the divergence can be calculated as the divergence of the
``shifted" expression
\begin{equation}
\int\int_{w_1>w_2} \; d w_{1} 
\left(\partial_{w_1}\phi_1 - {1\over R}n_{w_1}\phi_1\right) \;\;
d w_{2} \; \partial_{w_2}\phi_2 \,,
\end{equation}
in the free theory. This is just a convenient way of representing
the contraction of $\partial\phi_1$ with the interaction vertex 
$\int n \phi_1\bar{\partial}\phi_2$ in the action.
Therefore the divergence is equal to:
\begin{equation}
{1\over R}\log\epsilon \int dw \; n_w \,.
\end{equation}

\vspace{10pt}
\noindent
\underline{\bf 2. $(+-)\to (+)$.} This is is very similar:
\begin{equation}
\int\int_{\tau_1>\tau_2}
d w_1 \partial_{w_1}\phi_1 \;\; d \bar{w}_{2} \partial_{\bar{w}_2}\phi_2 \,.
\end{equation}
The divergence is equal to:
\begin{equation}
{1\over R}\log\epsilon \int dw \; n_w \,.
\end{equation}

\noindent
\underline{\bf 3.  $(-+)\to (+)$.}
\begin{equation}
\int\int_{\tau_1>\tau_2}
d \bar{w}_{1}\; \partial_{\bar{w}_1}\phi_1 \;\;\;
d w_{2} \;\partial_{{w}_2}\phi_2 \,.
\end{equation}
The divergence is the same as in $(+-)\to (+)$:
\begin{equation}
{1\over R}\log\epsilon \int dw \; n_w \,.
\end{equation}

\noindent
\underline{\bf 4.  $(--)\to (+)$.} 
Consider the integral with two $d\tau^-$:
\begin{equation}
\int\int_{w_1>w_2}d \bar{w}_{1}\;
\partial_{\bar{w}_1}\phi_1  \;\;\;
d \bar{w}_{2} \;\partial_{\bar{w}_2}\phi_2 \,.
\end{equation}
We were slightly surprized to find that 
this integral has a divergence proportional
to $\int d\tau^+ n_+$.
To calculate this divergence we have to evaluate the integral
over the position of the interaction vertex, with two contractions:
\begin{eqnarray}
\label{MinusMinusToPlus}
&& \int\int_{w_1 > w_2} \left(-{1\over \pi}\right)
\int d^2 v  n_w(v)
{\partial\over\partial \bar{w}_1}
{\partial\over\partial \bar{v} }
\log |w_1-v|^2 
{\partial\over\partial \bar{w}_2}
\log |w_2-v|^2
=\\
&& 
= \int d\bar{w}_1 d\bar{w}_2 
{w_1 - w_2\over (\bar{w}_1 - \bar{w}_2)^2}\; n_w =
- {1\over R} \log\epsilon \int dw\; n_w \,.
\end{eqnarray}

\subsubsection{Effect of the interaction on composite
currents like $[x , \partial x]$ }
\label{EffectOfInteractionOnComposite}
{\em Example when the interaction vertex  \underline{does not} lead
to a logarithmic divergence.} 
Consider the contour integral of the ``composite" operator:
\begin{equation}
\int d\bar{w}\; \phi_2\partial_{\bar{w}}\phi_1 \,.
\end{equation}
The integral over the interaction vertex becomes:
\begin{equation}
\int d^2 v {1\over (\bar{v} - \bar{w})^2}\log |v-\epsilon|^2 \,.
\end{equation}
This is convergent.
\vspace{10pt}

\noindent
{\em Example when the integration vertex \underline{does} lead to
a log divergence:}
\begin{equation}
\int dw \;\phi_2\partial_w\phi_1
\end{equation}

\vspace{10pt}
\begin{centering}
	\hfill
	\epsfxsize=2in
      \epsffile{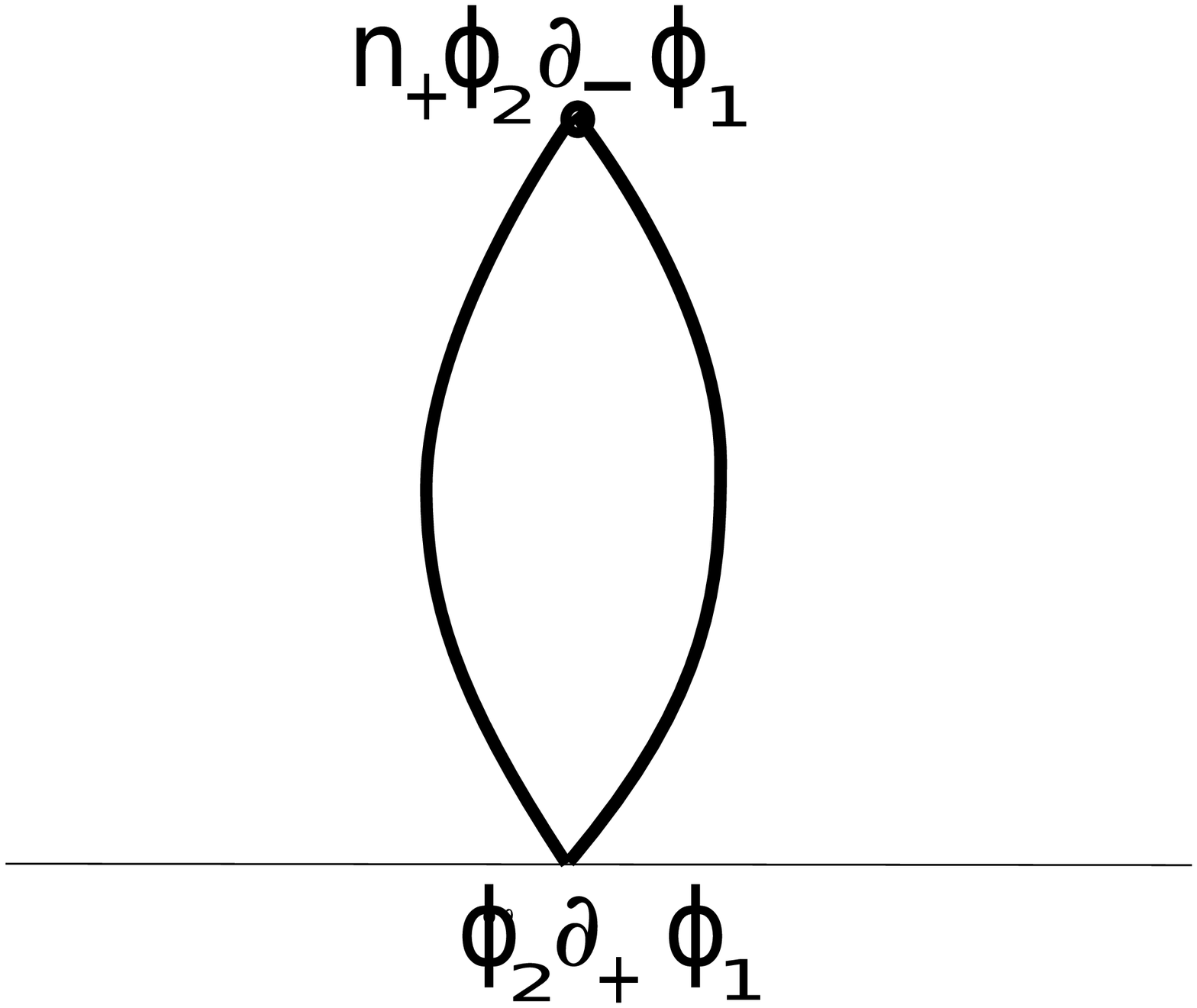}
      \hfill
\end{centering}
\vspace{10pt}

 The log divergence can be calculated by the following
trick. Replace 
\begin{equation}
\phi_2\partial_w\phi_1 \to 
\phi_2 \left(\partial_w\phi_1 - {1\over R} n_w \phi_1\right) \,,
\end{equation}
and calculate the OPE's as in the free field theory.
The logarithmic divergence is:
\begin{equation}
{1\over R}\log|\epsilon|^2 \int dw\; n_w \,.
\end{equation}
Therefore:
\begin{equation}
\int dw \; \phi_1\partial_w \phi_2 \mapsto
-{1\over R}\log|\epsilon|^2 \int dw \; n_w \,.
\end{equation}

\subsection{Algebraicity}
Notice that 1-loop logarithmic divergences are always of the
form $\int \Phi$ where $\Phi$ is a one-form composed of
the elementary fields and an element of $\psu(2,2|4)$. For example
we get the expressions like 
$\int \log\epsilon \; d\tau^+ [\vartheta,\partial_+ x]$
but we never get something like 
$\int \log\epsilon \; d\tau^+ \{\vartheta,\partial_+ x\}$;
the difference is that the anticommutator $\{\vartheta,\partial_+ x\}$ would not
belong to the Lie algebra $\psu(2,2|4)$.

\section{Calculation of logarithmic divergences}
\label{sec:CalculationOfLogDiv}

In this section we will apply the technique developed in Section
\ref{sec:RenormalizationWilsonLine}
and calculate the one loop logarithmic divergences in the
transfer matrix.
We will classify the divergences according to their dependence on the
field and the power of the spectral parameter $z$.

\subsection{Divergences proportional to $J_{2+}$}
\subsubsection{Coefficient of $z^{-6}$}
\label{J2PlusZMinus6}
There are two contributions. One comes from the double collision
of $J^{\dot{\alpha}}_{1+}(w_a)J_{1+}^{\dot{\beta}}(w_b)$:
\begin{equation}
{1\over z^6}{1\over 2}{1\over R^3} \log\epsilon
\partial x^{\mu} C^{\dot{\alpha}\alpha} 
\{t^1_{\dot{\alpha}} , [t_{\alpha}^3, t_{\mu}^2 ] \}
={1\over z^6}{1\over 4}{1\over R^3} \log \epsilon\; C_{odd}.\partial x \,.
\label{ZMinusSixDouble}
\end{equation}
The other one comes from the triple collisions, as described in
Section \ref{sec:TripleCollisions}:
\begin{equation}
-{1\over z^6}
{1\over 2}{1\over R^3} \log \epsilon\; (C_{odd}+C_{\bar{2}}).\partial x \,.
\label{ZMinusSixTriple}
\end{equation}
These two contributions cancel because of (\ref{CoddAndC2}).
Notice that this cancellation requires an interaction term
$-{1\over 2}{1\over R} f_{\mu\alpha\beta}\partial_+ x^{\mu}
\vartheta_L^{\alpha} \partial_-\vartheta_L^{\beta}$ in the
action. If this term were zero, for example, we would have
a coefficient $2$ in (\ref{OPEJ1PlusJ1Plus}), as in
(\ref{OPEJ3PlusJ3Plus}), and this would destroy the balance.

\subsubsection{Coefficient of $z^{-2}$}
\label{sec:J2PlusZm2}
As we have seen in Section \ref{sec:FieldRenormalization} there
are no internal log divergences in $J_{2+}$, at the order $R^{-3}$.
Therefore we are left with the contributions from:

\vspace{10pt}

\noindent
\underline{\em Double collision $J_{0+}J_{2+}$}:
\begin{equation}
{1\over 2} z^{-2} \log\epsilon \; C_{\bar{2}}.\partial_+ x \,.
\end{equation}
\underline{\em Double collision $J_{3+}J_{3+}$}:
\begin{equation}
{1\over 2}z^{-2}\log\epsilon \; C_{odd}.\partial_+ x \,.
\end{equation}
\underline{\em Double collision $J_{1+}J_{1-}$} does not
contribute.
\vspace{10pt}

\noindent
We conclude that \underline{ the log divergence proportional to 
$z^{-2} \partial_+ x$ is}:
\begin{equation}
\label{DivDxZm2}
-{1\over 2} z^{-2} \log\epsilon \; C_{\bar{2}}.\partial_+ x \,.
\end{equation}
It looks like we got a nonzero log divergence, but it turns
out that this divergence combines with the divergence 
proportional to $z^{-2} \partial_- x$ to a total
derivative. We will explain this 
in Section \ref{sec:DivergenceTotalDerivative}.

\subsubsection{Coefficient of $z^2$}
\label{sec:J2PlusZ2}
There are contributions from the triple collisions
$J_{1-}J_{3-}J_{2+}$ and $J_{2-}J_{2-}J_{2+}$ which are
proportional both to $z^2$.
This should cancel against the double collision $J_{0+}J_{2-}$ 
and $J_{3-}J_{3+}$,
which are also proportional to $z^2$.
Let us verify this. 
The contributions from the triple collisions are as usual:
\begin{eqnarray}
&&-{1\over 2R^3} z^2 \log\epsilon \left(
C^{\dot{\alpha}\alpha}
\{t^1_{\dot{\alpha}} , [t^3_{\alpha} , t^2_{\mu} ]\}
+
C^{\alpha\dot{\alpha}}
\{t^3_{\alpha} , [t^1_{\dot{\alpha}} , t^2_{\mu} ]\}
+C^{\kappa\lambda}[ t^2_{\kappa} , [t^2_{\lambda}, t^2_{\mu}] ]
\right)\; \partial_+x^{\mu}
= \nonumber \\
&&=-{1\over R^3}{1\over 2} z^2 \log\epsilon \;
(C_{odd}+C_{\bar{2}}).\partial_+x \,.
\label{TripleToRightJ2Plus}
\end{eqnarray}
Now let us evaluate the contribution of $J_{0+} J_{2-}$.
We get:
\begin{equation}
J_{0+}^{[\mu\nu]}(w_0) J_{2-}^{\lambda}(w_2)  = 
- {1\over 2 R^3}  
{\partial_+ x^{\kappa}(w_0) \over (\bar{w}_0 - \bar{w}_2)}
\fduu{\!\!\!\!\!\kappa}{\lambda}{[\mu\nu]} \,.
\end{equation}
This gives the logarithmic contribution:
\begin{equation}\label{DoubleJ0plusJ2minus}
{1\over  R^3} {1\over 2} z^2 \log\epsilon \; C_{\bar{2}}. \partial_+ x \,.
\end{equation}
There is also a $z^2$ contribution from $J_{3-}J_{3+}$:
\begin{eqnarray}
J_{3-}^{\beta}(w_b) J_{3+}^{\alpha}(w_a) & = &
	-{1\over R^3} {1 \over \bar{w}_b -\bar{w}_a }\partial_+ x^{\nu} 
	\fduu{\nu}{\beta}{\alpha} \,.
\end{eqnarray}
This leads to the divergence
\begin{equation}\label{J3minusJ3plusToRightJ2Plus}
{1\over R^3}{1\over 2} z^2 \log\epsilon\; C_{odd}. \partial_+ x \,.
\end{equation}
One can see that (\ref{TripleToRightJ2Plus}) + 
(\ref{DoubleJ0plusJ2minus}) + (\ref{J3minusJ3plusToRightJ2Plus}) = 0.
But there is also a contribution from the double collision 
$J_{1-}J_{1-}$, because of (\ref{MinusMinusToPlus}). 
Because of the interaction the OPE of $J_{1-}J_{1-}$ contains
$\partial_+ x$:
\begin{equation}
J_{1-}^{\dot{\alpha}} (w_a) 
J_{1-}^{\dot{\beta}}  (w_b)
= -\partial x^{\mu} \fduu{\mu}{\dot{\alpha}}{\dot{\beta}}
{w_a- w_b\over (\bar{w}_a - \bar{w}_b)^2} + \ldots \,.
\end{equation}
Therefore as explained in Section \ref{sec:EffectOnDoubleCollisions} 
double collision $J_{1-}J_{1-}$ contributes:
\begin{equation}
{1\over 4} z^2 \log\epsilon \; C_{odd}.\partial_+ x \;  d\tau^+ \,.
\end{equation}
Therefore the total \underline{\em log divergence proportional to 
$z^2 \partial_+ x$ is}:
\begin{equation}\label{DivDxZ2}
-{1\over 2} z^2 \log\epsilon\; C_{\bar{2}}.\partial_+ x \;  d\tau^+ \,.
\end{equation}

\subsubsection{ When the log divergence is a total
derivative.}
\label{sec:DivergenceTotalDerivative}
Logarithmic divergences  with the coefficient $z^{-2}$
apparently start at the order $R^{-3}$. But in fact they are total
derivatives and therefore contribute only through contact terms,
as we discussed in Section \ref{sec:DoubleCollisionsFieldDependent}.
Indeed, Eq. (\ref{DivDxZ2}) and the symmetry
 $(+\leftrightarrow -), (z \leftrightarrow z^{-1})$ implies that there
is the divergence:
\begin{equation}
-{1\over 2} z^{-2} \log\epsilon \;C_{\bar{2}}.\partial_- x \;  d\tau^- \,,
\end{equation}
from the double collision $J_{3+}J_{3+}$.
This and (\ref{DivDxZm2}) implies that the total divergence
proportional to $z^{-2}$ at the order $R^{-3}$ is the total derivative:
\begin{equation}\label{Zm2TotalDerivative}
-{1\over 2} z^{-2}
\log\epsilon \;C_{\bar{2}}.dx \,.
\end{equation}
Similarly the divergent term (\ref{DivDxZ2}) and the order $z^2$
contribution to the divergence proportional to $\partial_- x$
gives
\begin{equation}
-{1\over 2} z^2 \log\epsilon\; C_{\bar{2}}. dx \,.
\end{equation}
Integration of (\ref{Zm2TotalDerivative})
gives us the boundary terms. Some of these boundary terms are of the form:
\begin{equation}\label{BoundaryTerms}
{1\over R^4}  \log\epsilon  \left(\hspace{9pt}
-{1\over 2 z^3} [\partial_+ \vartheta_L, C_{\bar{2}}.x] 
\hspace{9pt} \mbox{or} \hspace{9pt} 
-{1\over 2 z^4} [\partial_+ x , C_{\bar{2}}.x] 
\hspace{9pt} \mbox{or} \hspace{9pt}
-{1\over 2 z^5} [\partial_+ \vartheta_R, C_{\bar{2}}.x]
\hspace{9pt} \right) \,.
\end{equation}
These terms are of the order $R^{-4}$ and of the same structure
as the logarithmic divergences arizing from the double collisions
and internal divergences in the order $R^{-4}$. 
For example, consider the term 
\begin{equation}\label{BoundaryTermThetaRx}
-{1\over R^4}{1\over 2 z^5} \log\epsilon [\partial_+ \vartheta_R, x] \,.
\end{equation}
A divergent term of the same structure appears in the double collision
$J_{1+} J_{2+}$. Indeed, there is a term in the action of the form
${1\over R^2} \partial_+\vartheta_R\partial_-x\vartheta_L x$.
This term leads to the divergence of the form (\ref{BoundaryTermThetaRx})
from the double collision 
${1\over R^2} \partial_+\vartheta_R(w) \partial_+ x(0)$, $w\to 0$.
Another possibility for the boundary terms is:
\begin{equation}
\label{BoundaryTerms2}
{1\over R^4}  \log\epsilon  \left(\hspace{10pt}
-{1\over 2} z [\partial_- \vartheta_L, x] 
\hspace{20pt} \mbox{or} \hspace{20pt} 
-{1\over 2}[\partial_- x , x] 
\hspace{20pt} \mbox{or} \hspace{20pt}
-{1\over 2z}[\partial_- \vartheta_R, x]
\hspace{10pt} \right) \,.
\end{equation}
They also interfere with various double collisions, just like
the terms in (\ref{BoundaryTerms}).
There are also the boundary terms  of the form
\begin{equation}
-{1\over R^5}  \log\epsilon {1\over 2 z^2} 
\left( 1 -{1\over z^4} \right)
[x,N_+] \,.
\end{equation}
They ``interfere" with the logarithmic divergence which appears
in the double collision $J_{2+}N_+$, because of the nonlinear
term $J_{0-}N_+$ in the action.

In Section (\ref{sec:DivXX}) we will explicitly verify the cancellation
of the contact terms from integrating the total derivative $dx$ in
$\int {1\over z^2}\log\epsilon \; dx \int {1\over z^2} d\tau^+\partial x$ 
against other divergences proportional to $z^{-4}[x,\partial_+ x]$.

We see that the left ${\mathfrak{g}}_{\bar{2}}$ divergences with the
coefficient $z^{-2}$ actually belong to the order $R^{-4}$, rather
than the order $R^{-3}$.
Integrating the total derivative can be understood as a 
 $z$-dependent
gauge transformation of $J(z)$, see Section \ref{sec:GaugeTransformation}.

\subsection{Divergences proportional to $J_{3+}$ and $J_{1+}$}
\label{sec:J3PlusAndJ1Plus}
The divergences proportional to $J_{3+}$ and $J_{1+}$ at the order $R^{-3}$
are zero because of the identity (\ref{NoSpinorDiv}).
Let us explain this for $J_{3+}$.
Coefficient of $z^{-1}$ could come from the following sources:
\begin{itemize}

\item	``internal" anomalous dimension of $J_{3+}$ from the interaction
	term \\
	$-{1\over 2}{1\over R} f_{\mu\dot{\alpha}\dot{\beta}}\partial_- x^{\mu}
	\vartheta_R^{\dot{\alpha}} \partial_+\vartheta_R^{\dot{\beta}}$ in the
	action

\item	the double collision $J_{0+}J_{3+}$

\item	double collisions $J_{1+}J_{2-}$ and $J_{2+}J_{1-}$, because
	of the interaction.

\end{itemize}
All these contributions are zero, being proportional to
one of the expressions in (\ref{NoSpinorDiv}). 
The coefficient of $z^{-5}$ is zero for the same reason.
There are the following potential contributions:
\begin{itemize}

\item	double collisions $J_{1+}J_{2+}$ 

\item	triple collisions,

\end{itemize}
but they are all zero because of (\ref{NoSpinorDiv}).
There could be also terms proportional to $z^3$, from the 
 double collisions $J_{0+}J_{3-}$ and from the triple collisions 
of the ``wrapping" type $J_-J_{3+}J_-$; they are zero for the 
same reason.

\subsection{Divergences proportional to $N_+$.}
\label{sec:DivergencesProportionalToN}

\subsubsection{Ghosts}
The terms in the action containing ghosts are
\begin{equation}
S_{ghosts}= 
{1\over \pi}\int d^2 v \;
\mbox{Str}\left( w_{1+}(\partial_- \lambda_3 +  [J_{0-},\lambda_3]) 
+ \tilde{w}_{3-}(\partial_+ \tilde{\lambda}_1 + [J_{0+},\tilde{\lambda}_1])
- N_+N_- \right)
\end{equation}
The ghost current is defined as
\begin{equation}
N_+^{[\mu\nu]} = -{1\over R^2} \{ w_+ , \lambda \} =
-{1\over R^2} w_+^{\dot{\alpha}} \lambda^{\beta} 
\fddu{\!\!\!\!\!\dot{\alpha}}{\beta}{[\mu\nu]} \,,
\end{equation}
and has OPE
\begin{equation}
N_+^{[\mu_1\nu_1]}(v) N_+^{[\mu_2\nu_2]}(0)
={1\over R^2}{1\over v} 
\fduu{\hspace{-30pt}[\mu_3\nu_3]}{\hspace{17pt}[\mu_1\nu_1]}{[\mu_2\nu_2]}
N_+^{[\mu_3\nu_3]}+{1\over R^4}{c\over v^2} + \ldots \,,
\end{equation}
where $c$ is a c-number. This c-number would play a role in
the logarithmic divergences, but at higher orders.

\subsubsection{Logarithmic divergence proportional to $N_+$.}
\label{sec:DivergenceNPlus}
Let us consider the renormalization of the coefficient of $N_+$.
The potentially divergent expressions arise in the order $R^{-4}$.
For the expressions proportional to the matter fields, we
verified the cancellation of the logarithmic divergences
up at the order $R^{-3}$. But for the expressions contating
ghosts we will calculate all the potentially divergent 
terms in the order $R^{-4}$.
There are the following sources of the logarithmic
divergence proportional to $N_+$:
\vspace{10pt}

\noindent
\underline{\em  Double collisions $N_+N_+$.}
The first source of the anomalous dimension is the $N_+N_+$ collision.
The corresponding contribution to the anomalous dimension is:
\begin{equation}
{1\over 2} {1\over R^2}\left( 1- {1\over z^4}\right)^2
 \log\epsilon \; C_0.N_+ \,.
\label{DoubleNNtoN}
\end{equation}

\noindent
\underline{\em Triple collisions $J_+N_+J_+$.}
The second source of the logarithmic divergence 
is the ``wrapping" of $t^a \otimes t^a$ and
$t^{\alpha}\otimes t_{\alpha}$ and $t_{\alpha} \otimes t^{\alpha}$
around $t_{\mu\nu}$.  It is proportional to
\begin{eqnarray}
-{1\over 2}{1\over R^2}{1\over z^4} \left(1-{1\over z^4}\right) \log\epsilon 
(C-C_0). N_+ \,.
\label{WrappingNLeft}
\end{eqnarray}
\vspace{10pt}

\noindent
\underline{\em Mixing with $J_{0+}$.}
Another contribution comes from the 
mixing of $J_{0+}$ into $N_+$
caused by the term $\mbox{str}\;N_+J_{0-}$ in the
action, as described in Section \ref{EffectOfInteractionOnComposite}:  
\begin{equation}\label{LeftInternalJ0ToN}
-J_{0+}\to {1\over 2}{1\over R^2}
\log |\epsilon|^2 (C-C_0). N_+ \,.
\end{equation}
\vspace{10pt}

\noindent
\underline{\em  Double collisions
$J_{1+}J_{3+}$ and $J_{2+}J_{2+}$.} For example, the double collision
$J_{2+}J_{2+}$ leads to the log divergence $\simeq N_+$ which can be
effectively described as the log divergence of this collision:
\[
\left(\partial_+x + {1\over 2} [N_+,x]\right)  \llra
\left(\partial_+x + {1\over 2} [N_+,x]\right)  \,.
\]
The total contribution
from the double collision $J_{2+}J_{2+}$ and $J_{1+}J_{3+}$ is:
\begin{equation}\label{DoubleCollisionToNLeft}
-{1\over 2}{1\over R^2}{1\over z^4} \log\epsilon \; (C-C_0). N_+ \,.
\end{equation}

\noindent
\underline{\em Triple collision $J_{-}N_+J_{-}$.}
These triple collisions contribute:
\begin{eqnarray}
-{1\over 2}{1\over R^2} z^4 \left(1-{1\over z^4}\right) \log\epsilon 
(C-C_0). N_+ \,.
\label{WrappingNRight}
\end{eqnarray}
\vspace{10pt}

\noindent
\underline{\em Mixing with $J_{0-}$.} There is no such mixing,
see Section \ref{EffectOfInteractionOnComposite}.
\vspace{10pt}

\noindent
\underline{\em Double collisions $J_- J_+$.} Their contribution 
can be effectively calculated by evaluating in the free theory
the $N_+$-singularity in the  $J_{2-}J_{2+}$ collision:
\begin{equation}
\left( \partial_- x + [ N_- , x] \right)
\llra
\left( \partial_+ x + [ N_+ , x] \right) \,,
\end{equation}
and similar $J_{1-}J_{3+}$ and $J_{3-}J_{1+}$ collisions.
The result is
\begin{equation}\label{JMinusJPlusToN}
-{1\over R^2} \log\epsilon (C-C_0).N_+ \,.
\end{equation}
\vspace{10pt}

\noindent
\underline{\em Double collisions $J_- J_-$.} 
These collisions contribute because of (\ref{MinusMinusToPlus}):
\begin{equation}\label{JMinusJMinusToN}
{1\over 2}{1\over R^2} 
z^4\log\epsilon (C-C_0).N_+ \,.
\end{equation}
The total result is that the logarithmic divergences
proportional to $N_+$ add up to zero:\\
(\ref{DoubleNNtoN}) + (\ref{WrappingNLeft}) +
(\ref{LeftInternalJ0ToN}) + (\ref{DoubleCollisionToNLeft}) +
(\ref{WrappingNRight}) + (\ref{JMinusJPlusToN}) +
(\ref{JMinusJMinusToN}) = 0

\subsection{Divergences of the type $x\; \partial_+ x$}
\label{sec:DivXX}
\subsubsection{The coefficient of $z^{-4}x\;\partial_+x$}
\label{sec:DivXXzMinus4}
First let us calculate the coefficient of ${1\over z^4} [\partial_+ x , x]$.
Notice that in the expansion of the transfer matrix we get the term
$\int \left(-{1\over 2} [dx , x]\right)$ which is proportional to $z^0$, 
but we do not
have classically any terms which would be proportional to $z^4$ or $z^{-4}$.
There are the following divergent contributions:

\vspace{10pt}

\noindent
\underline{\em Triple collisions.}  There are triple collisions of the
form:
\[
J_{2+} \llra J_{2+} \llra \left( -{1\over 2} [\partial_+ x , x] \right)
\;\;\;\mbox{and}\;\;\; 
J_{3+} \llra J_{1+} \llra \left( -{1\over 2} [\partial_+ x , x] \right)
\]
As we discussed in Section \ref{sec:TripleCollisions} the contribution
of the triple collisions is due to the second order terms which appear
in the OPE of $\partial_+x$ with $\partial_+ x$, or in the OPE of
$\partial_+\vartheta_R$ with $\partial_+ \vartheta_L$. 
Let us first consider the triple collisions with two $J_{2+}$.
One contribution comes from the contraction of $\partial_+ x$ 
in two $J_{2+}$; the divergence is:
\begin{equation}
{1\over 4}\log\epsilon\; C^{\mu\nu} 
[t^2_{\mu} , [t^2_{\nu} , [\partial_+x , x] ]] \,.
\end{equation}
The other contribution comes from the contraction of 
$\partial_+ x $ in $J_{2+}$ with $\partial_+ x$ in $[\partial_+ x, x]$,
this gives the following divergence:
\begin{equation}
{1\over 2}\log\epsilon\; 
C^{\mu\nu} [[t^2_{\mu} , x ]  , [t^2_{\nu}, \partial_+ x ] ] \,.
\end{equation}
The total of contributions from 
$J_{2+}\llra J_{2+}\llra \left(-{1\over 2}[\partial_+x , x]\right)$ 
collisions is:
\begin{equation}\label{J2plusJ2plusToXX}
{1\over 2} \log\epsilon [\partial_+x , C_{\bar{2}}.x] \,.
\end{equation}
Now let us consider the collision with $J_{1+}$ and $J_{3+}$.
The result is:
\begin{equation}\label{J1plusJ3plusToXX}
{1\over 4} \log\epsilon 
\left(
C^{\alpha\dot{\alpha}} \{ t^3_{\alpha} , [ t^1_{\dot{\alpha}} ,
[\partial_+ x , x]]\}+
C^{\dot{\alpha}\alpha} \{ t^1_{\dot{\alpha}} , [t^3_{\alpha} ,
[\partial_+ x , x]]\}
\right)
=-{1\over 2}\log\epsilon [ \partial_+ x , C_{\bar{2}}.x ]   \,.
\end{equation}
Therefore {\em the total contribution from the triple collisions} is
zero: (\ref{J2plusJ2plusToXX}) + (\ref{J1plusJ3plusToXX}) = 0.

\vspace{10pt}
\noindent \underline{\em Double collisions.}
One possible double collision is $J_{2+}\llra J_{2+}$. But in fact this
double collision does not contribute to $[\partial_+ x , x]$.
(Let us prove that $J_{2+}J_{2+}$ does not contribute. The terms in $J_{2+}$
which could contribute are 
$-\partial_+ x - {1\over 6} [x, [x, \partial_+x]]$.
There is a contribution from the collision 
$\partial_+ x \llra {1\over 6} [x, [x, \partial_+x]]$ but it cancels
with the contribution from the collision $\partial_+ x \llra \partial_+ x$
which arises because there is the term 
$-{1\over 6} [\partial_+ x,x][\partial_-x,x]$ in the action.)

But there is another double collision $J_{1+}\llra J_{3+}$, and it does
give a nonzero contribution.
Let us calculate the contribution of $J_{1+}\llra J_{3+}$ to the log
divergence proportional to $[\partial_+ x ,x]$.
The relevant terms in the expansion of the currents are:
\begin{eqnarray}
-J_{1+} & = & \partial_+\vartheta_R - 
		{1\over 2} [\vartheta_R , [\partial_+ x , x]] +\ldots
\\
-J_{3+} & = & \partial_+\vartheta_L -
		{1\over 2} [\vartheta_L , [\partial_+ x , x]] +\ldots \,.
\end{eqnarray}
The relevant terms in the action are:
\begin{equation}
- {1\over 8}  [\vartheta_L, \partial_- \vartheta_R] [x, \partial_+ x]
- {3\over 8}  [\vartheta_R, \partial_- \vartheta_L] [x, \partial_+ x] \,.
\end{equation}
This means that the divergence is the same as if we collided 
\[\left(\partial_+\vartheta_R +
\left(-{1\over 2} +{3\over 8}\right) 
[\vartheta_R , [\partial_+ x , x]] \right)
\llra
\left(\partial_+\vartheta_L +
\left(-{1\over 2} +{1\over 8}\right) 
[\vartheta_L , [\partial_+ x , x]] \right) \,,
\]
in the free theory. This gives the \underline{\em contribution
from double collisions}:
\begin{equation}
-{1\over 4} \log\epsilon\; C_{odd}.[\partial_+x , x] =
 {1\over 2} \log\epsilon\; [\partial_+ x , C_{\bar{2}}.x] \,.
\end{equation}

\vspace{10pt}
\noindent
\underline{\em Contribution from the total derivative.}
There is a contribution from (\ref{BoundaryTerms}):
\begin{equation}
-{1\over 2} \log\epsilon\; [\partial_+ x , C_{\bar{2}}.x] \,.
\end{equation}
We see that the contribution from the boundary terms cancels
the contribution from the double collisions, and therefore
the total log divergence of the type $z^{-4}x\; \partial_+ x$ is zero.

\subsubsection{Coefficient of $z^0 x\;\partial_+ x$}
We did not calculate this coefficient.
But we have seen that the coefficient to $z^{-4}x\; \partial_+ x$ is zero,
and we will see that the coefficient of $z^4 x\;\partial_+x$ is also zero.
We know there should not be any log divergence at $z=1$.
Therefore the log divergence proportional to $z^0 x\;\partial_+ x$
should be zero.

\subsubsection{Coefficient of $z^4$}
\underline{\em Contribution from triple collisions.}
\label{sec:DivXXzPlus4}
Triple collisions of the type $J_{1-}J_{3-} J_{0+}$ and
$J_{2-} J_{2-} J_{0+}$ contribute to the divergence of the
form $z^4 [\partial_+ x , x]$ and their contribution is equal to:
\begin{equation}\label{TripleToZ4XX}
{1\over 4}\log\epsilon (C_{odd}+C_{\bar{2}}).[\partial_+ x , x]=
-{3\over 8} \log\epsilon [\partial_+ x , C_{\bar{2}}.x] \,.
\end{equation}
\underline{\em Contribution from double collisions.}
There is a contribution from $J_{2-}J_{2-}$ and a contribution
from $J_{1-}J_{3-}$. Let us first consider the contribution from 
$J_{2-}J_{2-}$. The relevant interaction term in the action is
$-{1\over 6} [\partial_+ x , x] [\partial_- x , x ]$.
To get
the divergence of the form $[\partial_+ x , x]$ we 
should have $\partial_- x$ in the interaction vertex contracted with
 a $\partial_- x$ in one of the $J_{2-}$. We get:
\begin{eqnarray}
&& -{1\over 6} {1\over \pi} \int d^2v 
{1\over (\bar{w}_L - \bar{v})^2 (\bar{w}_R-\bar{v})}
\left(
\mbox{str} ( t_{\nu}^2 [ t_{\mu}^2 , [x,\partial_+ x]]) +
\mbox{str} ( t_{\nu}^2 [ x , [t_{\mu}^2 ,\partial_+ x]])\right)
[t_{\nu}^2 , t_{\mu}^2] 
=\nonumber \\ 
&& = -{1\over 6} \log\epsilon \left( 
[[t_{\mu}^2 , [x,\partial_+ x]] , t_{\mu}^2] +
[[x, [t_{\mu}^2 ,\partial_+ x]] , t_{\mu}^2] \right)
=\nonumber \\
&& = -{1\over 8} \log\epsilon [\partial_+ x , C_{\bar{2}}.x]\,.
\label{J2J2ToZ4XX}
\end{eqnarray}
Now let us consider the double collision $J_{1-}J_{3-}$. The
relevant interaction vertices are 
\[
- {1\over 8}  [\vartheta_L, \partial_- \vartheta_R] [x, \partial_+ x]
- {3\over 8}  [\vartheta_R, \partial_- \vartheta_L] [x, \partial_+ x]\,.
\]
These two vertices give the same contribution and add up to:
\begin{eqnarray}
&& -{1\over 2} {1\over\pi} \int d^2 v
{1\over (\bar{w}_L - \bar{v})^2 (\bar{w}_R-\bar{v})}
C^{\alpha\dot{\beta}} C^{\gamma\dot{\alpha}}
\mbox{str}(t^1_{\dot{\beta}}[t^3_{\gamma}, [x , \partial_+ x]])
 \{t^3_{\alpha},t^1_{\dot{\alpha}}\}
=\nonumber\\
&& = -{1\over 4} \log\epsilon C_{odd}.[\partial_+ x , x]
= {1\over 2} \log\epsilon [\partial_+ x , C_{\bar{2}}.x] \,.
\label{J1J3ToZ4XX}
\end{eqnarray}
The total contribution 
(\ref{TripleToZ4XX}) + (\ref{J2J2ToZ4XX}) + (\ref{J1J3ToZ4XX}) = 0.
Therefore there is no logarithmic divergence of the type 
$z^4 [\partial_+ x , x]$.

\subsection{Bulk divergences and divergences associated to the boundary}
\label{sec:GaugeTransformation}
In this section we have collected evidence that the
logarithmic divergences of the transfer matrix at one loop
are zero modulo the total derivative. The total derivative
was described in Section \ref{sec:DivergenceTotalDerivative}.
This suggests that the logarthmic divergences of the transfer
matrix have the following form:
\begin{equation}
\Omega(z)=f(\epsilon,z) \Omega(z)^{finite} f(\epsilon,z)^{-1}\,.
\end{equation}
where $f(z,\epsilon)$ is a $z$-dependent gauge transformation:
\begin{equation}\label{ZDependentGaugeTransformation}
f(z,\epsilon)=\exp\left( 
-{1\over 2R^2}\left(z^2+{1\over z^2}\right)\log\epsilon \;
C_{\bar{2}}.x + \ldots\right)\,.
\end{equation}
Dots denote terms of the higher power in $1\over R$.
In this sense, we can say that 
the divergences are absorbed in a $z$-dependent gauge
transformation. But notice that when we compute the path ordered exponential
of $-\int_C J$ over an {\em open} contour $C$, we get additional divergences
associated to the endpoints. 

We want to investigate the
following question: can we distinguish between the bulk divergences,
which are total derivative ``propagating" to the endpoint,
and the boundary divergences which are ``inherent to the endpoint"?
Instead of considering open contour with endpoints, it is more convenient
to consider a closed contour and insert a constant matrix $X$ 
at some point $\tau_0=(\tau_0^+,\tau_0^-)$ inside the 
contour\footnote{Such an object is not gauge invariant, with respect to the
${\bf g}_{\bar{0}}$ gauge transformations. But let us 
fix the gauge as in (\ref{VielbeinGaugeFixing}) and consider
this expression in the fixed gauge, just as an example.}:
\begin{equation}
P\left[ X(\tau_0)\exp\left(-\int J(z)\right)\right]
\end{equation}
It seems that there are two different types of divergences associated with 
the insertion of $X$: 
\begin{enumerate}

\item 	the divergences of 
	$f^{-1}Xf$ which arise because $f$ is divergent, see 
	Eq. (\ref{ZDependentGaugeTransformation}); this is the
	effect of the divergences in the bulk of the contour,
	which are total derivatives and therefore ``propagate"
	to the insertion point

\item	the divergences of diagramms localized near
	the insertion of $X$, as in Section \ref{sec:TripleCollisions}

\end{enumerate}
But in fact the difference between these two types of divergences is a matter
of convention. Indeed, let us return to Section \ref{sec:J2PlusZm2}
and remember how we calculated the divergence proportional to
$z^{-2}\partial x$. One of the contributions to the divergence was
from the collision $J_{2+}\llra J_{0+}$ which in the leading order
was $-\partial_+ x(w_2) \llra {1\over 2} [\partial_+ x, x](w_0)$.
The ambiguity arises when we decide whether to first integrate over $w_2$
and then over $w_0$, or the other way around. 
We agreed in Section \ref{sec:TripleCollisionsFieldDependent} to integrate
first over $w_2$, and followed this prescription in Section \ref{sec:J2PlusZm2}.
It was more convenient because with this prescription only the first order pole
(from the contraction of $\partial_+x(w_2)$ with
$x(w_0)$) contributes to the log divergence.
If we integrated first over $w_0$, we would have a contribution to
the log divergence from the second order pole.
 But of course the divergences ``associated to 
the insertion of $X$" also depend on the order of integration.
For example, the collision $X\llra [\partial_+x , x] \llra \partial_+ x $
will not contribute to the log divergence ``of the insertion $X$"
if we first integrate over the position of $\partial_+x$, but will
contribute if we first integrate over the position of $[\partial_+x , x]$.

The lesson is that if we agreed on the order of integration in
the bulk of the contour, we should use it consistently also when
computing the log divergences associated to the boundary.
A different arrangement of the order of integrations will lead
to the different distribution of the log divergences between
the bulk total derivative terms and the collisions with the boundary.
In other words, when the contour is open, {\em there is no good distinction
between the log divergences which come from the total derivative divergences
in the bulk and the log divergences coming from the boundary effects}.

If the log divergences in the bulk of the contour are total derivatives,
this means that it is possible to choose a prescription for the order of
integrations such that the bulk divergence is zero. Of course, if we want
to compute for the open contour the anomalous dimension of the endpoints, 
or the anomalous dimension of some insertion, then we have to consistently follow
the same prescription calculating the boundary divergences. 


\section{Logarithmic divergences and global symmetries}
\label{sec:LogDivAndGlobalCharges}

Global symmetries described in Section \ref{sec:GlobalShifts} impose very strong
constraints on the divergences, and actually imply that the cancellation of the 
1-loop logarithmic divergences follows from the cancellation of the simplest
possible divergent expressions, those proportional to $\partial_{\pm} x$
and $\partial_{\pm}\vartheta$. 
This section consists of two parts. In the first part we will show that the global
symmetries together with the results of the previous section imply that 
the 1-loop logarithmic divergences vanish. In the second part we demonstrate the
consistency of the short distance singularities in the product $J_{2+} J_{2+}$
with the global symmetries. 

\subsection{Vanishing of the 1-loop logarithmic divergences}
Let us start with the log divergences proportional to $x\partial x$.
In the previous section we demonstrated by explicit calculations that there
are no such divergences. 
But in fact  the cancellation of this type of divergences automatically follows
from the cancellation of the log divergences of the form $\partial x$ and the
invariance under the global symmetries described in Section \ref{sec:GlobalShifts}.

In the previous section we have shown that the divergent terms in the bulk of the form
$\log\epsilon\; z^{4k-2}\partial_{\pm} x$ and 
$\log\epsilon\; z^{2k-1}\partial_{\pm}\vartheta$ are all total derivatives.
Let us make a $z$-dependent gauge transformation eliminating these total derivatives.
After such a $z$-dependent gauge transformation there are no log divergent
terms of the form $\log\epsilon\; z^{4k-2}\partial_{\pm} x$ and 
$\log\epsilon\; z^{2k-1}\partial_{\pm}\vartheta$.

Let us first prove that the cancellation of the one-loop divergences of the form 
$\log\epsilon \; z^{4k-2}\partial_+ x$ implies the cancellation of the one-loop divergences 
of the form $ \log\epsilon \;
z^{4k}(\alpha\; [x , \partial_+ x] d\tau^+ 
+ \beta\; [x , \partial_- x] d\tau^-)$.
Indeed the shift of this expression by $\xi$ would be
\begin{equation}\label{ShiftOfPotentialOneLoopJ0}
\log\epsilon \;
z^{4k}(\alpha\; [\xi, \partial_+ x] d\tau^+ 
+ \beta\; [\xi, \partial_- x] d\tau^-) \,,
\end{equation}
plus higher order terms\footnote{Note that $\alpha$ and $\beta$ are typically not
c-numbers, but contain Casimir operators $C_{\bar{0}}$ acting on $[x,\partial x]$.}.
This contradicts the shift invariance unless $\alpha=\beta$, in which
case (\ref{ShiftOfPotentialOneLoopJ0}) is 
the total derivative $z^{4k}\alpha [\xi, dx]$. But even if $\alpha=\beta$, the total
derivative being integrated by parts would hit $-\int z^{-2}J_{\bar{2}+}d\tau^+$ and 
give 
\begin{equation}\label{StructureOfVariation}
\int z^{4k-2}\log\epsilon\; [\partial_+x , \alpha [\xi, x]] d\tau^+ \,,
\end{equation}
which cannot be cancelled by anything. Indeed, we have shown that there are no counterterms
of the form $z^{4k-2}\log\epsilon\; \partial_+ x$. The possible counterterms
of the form $z^{4k-2}\log\epsilon\; [x,[x,\partial_+x]]$ would have the variation
\begin{equation}\label{VariationOfPossibleCounterterms}
z^{4k-2}\log\epsilon\; ( [\xi,[x,\partial_+x]] + [x,[\xi,\partial_+x]] ) \,,
\end{equation}
which has a different structure from (\ref{StructureOfVariation}).
In other words, the expression (\ref{StructureOfVariation}) can not be represented
as $S_{\xi}$ of something\footnote{But if there was a divergent term of the form
$z^{-2}\int\log\epsilon\; \partial_+x d\tau^+$, then the variation of such a term,
because of the higher order terms in (\ref{GlobalShiftOfX}), would be of the form
${1\over 3}z^{-2}\log\epsilon\;([\partial_+x, [x,\xi]]+[x,[\partial_+x,\xi]])$, and this
could combine with (\ref{VariationOfPossibleCounterterms}) to cancel 
(\ref{StructureOfVariation}).  } 
of the type $xx\partial_+x$.
This argument shows that there are no divergences of the form $z^{4k} [x,\partial_{\pm}x]$,
and also no divergences of the form $z^{4k-2} [x, [x,\partial_{\pm}x]]$.

Let us now rule out the divergences of the type $z^{2k}[\vartheta,\partial_{\pm}\vartheta]$.
 At the one loop level we could only have divergences
proportional to one of these expressions:
\begin{eqnarray}
& 
z^{\pm 8}[\vartheta_L,\partial_{\pm}\vartheta_R], \;
z^{\pm 8}[\vartheta_R,\partial_{\pm}\vartheta_L], \;
z^{\pm 6}[\vartheta_L,\partial_{\pm}\vartheta_L], \;
z^{\pm 6}[\vartheta_R,\partial_{\pm}\vartheta_R], \;
z^{\pm 4}[\vartheta_L,\partial_{\pm}\vartheta_R], \;
z^{\pm 4}[\vartheta_R,\partial_{\pm}\vartheta_L], \;
& \nonumber \\
& z^{\pm 2}[\vartheta_L,\partial_{\pm}\vartheta_L], \;
z^{\pm 2}[\vartheta_R,\partial_{\pm}\vartheta_R],  \;
[\vartheta_L,\partial_{\pm}\vartheta_R], \;
[\vartheta_R,\partial_{\pm}\vartheta_L]& \,.
\label{ListOfPossibleDivergences} 
\end{eqnarray} 
For example, there are no divergences of the form
$\log\epsilon\; z^{-10}[\vartheta_L,\partial_{\pm}\vartheta_L]$, 
because such divergences would require
colliding more than three currents; at the one loop level there are no log divergences
coming from the multiple collisions of the order higher than double and triple.
Also, there are no divergences of the form $z^{-9} [\vartheta_R, \partial_{\pm}x]$
or $z^{-9} [\partial_{\pm}\vartheta_R, x]$.
By counting the powers of $z$, such divergences could only appear in a triple collision 
$J_{1+}J_{1+}J_{1+}$, but
there are no suitable contractions. Therefore the potential divergent terms with the
highest negative power of $z$ are $z^{-8}[\vartheta_L,\partial_{\pm}\vartheta_R]$ and
$z^{-8}[\partial_{\pm}\vartheta_L,\vartheta_R]$.

We use the invariance under the shifts and the supershifts: 
\[
\delta x= \xi+\ldots\;\; , \;\;\;
\delta \vartheta_{L,R}=\zeta_{L,R} +\ldots \,.
\]
Let us first rule out the possible divergence with the highest negative power of $z$:
\[
\int\log\epsilon 
\left(
\alpha z^{-8}[\vartheta_L,\partial_+\vartheta_R]d\tau^+ +
\beta z^{-8}[\vartheta_L,\partial_-\vartheta_R]d\tau^- 
\right) \,.
\]
For the variation to be a total derivative it is necessary to have $\alpha=\beta$,
and then we get the variation $\int \log\epsilon \; z^{-8}[\zeta_L, d\vartheta_R]$.
Integrating this expression we would hit (for example) the classical term 
$\int d\tau^+ z^{-3}\partial_+\vartheta_R$ and get the contact term
$\int d\tau^+ z^{-11} [\partial_+\vartheta_R , [\zeta_L,\vartheta_R] ]$. 
Notice that this contact term could be cancelled by the variation of the divergent
term proportional to 
$z^{-11}[\partial_+ \vartheta_R, [\vartheta_L, \vartheta_R]]$ under $\delta_{\zeta_L}$,
if there is such a divergent term. But such a divergent term would also have a nonzero
variation under $\delta_{\zeta_R}$. 
The only way to match the variation under $\delta_{\zeta_R}$ is to have also the 
divergence proportional to $z^{-8}  [d\vartheta_L , \vartheta_R]$, which combines
together with $z^{-8}[\vartheta_L,d\vartheta_R]$ to the total
divergent term $z^{-8}d[\vartheta_L,\vartheta_R]$.
 This can be gauged away
by a $z$-dependent gauge transformation. 

Other possible divergences from the
list (\ref{ListOfPossibleDivergences}) and also divergences 
of the type $z^{2k+1}[x,\partial_{\pm}\vartheta]$ 
and $z^{2k+1}[\partial_{\pm}x, \vartheta]$
could be ruled out by essentially the same arguments, first those
proportional to $z^{-7}$, then $z^{-6}$, and so on.  Suppose that we have a divergent
term of the form $z^{-k}\log\epsilon \int Y$, where $Y$ is a 1-form  
quadratic in the elementary fields, for example $Y=[x,d\vartheta]$. We should have
$\delta Y=dZ$, and then the variation will give many contact terms including this one:
$z^{-k-3}[\partial_+\vartheta_R, Z]$. This should be cancelled by $\delta$ of
some divergent term cubic in the elementary fields, therefore we should have
$Z=\delta X$. We have $\delta(Y-dX)=0$ and this implies $Y=dX$ because 
$Y-dX$ is quadratic in elementary fields but contains only one derivative.

Therefore the near flat space expansion of the logarithmic divergences should start
with the terms  of the form $\int \log\epsilon \; \Phi(x,\theta)$
where $\Phi(x,\vartheta)$ is a worldsheet 
1-form composed of three or more
$x$ and $\vartheta$, for example $\Phi=z^{-2}[\vartheta_R,[\vartheta_L,*dx]]$. 
Invariance of the log divergences under shifts requires that $\delta\Phi=d \Psi$,
where $\Psi$ is some expression composed of at least two $\vartheta$ or $x$ and
one $\delta\vartheta$ or $\delta x$.
It turns out that this implies\footnote{This is because $\delta\Phi=d \Psi$ 
implies $\delta d\Phi=0$, and
since $d\Phi$ is a 2-form composed of at least three $x$ and $\vartheta$
this implies that $d\Phi=0$. And $d\Phi=0$ implies that $\Phi=dG$. Indeed, 
$\Phi$ contains at least three elementary fields, {\it e.g.} $x\vartheta *d x$, 
To the leading order in $R^{-2}$ we can think of $\Phi$ as the charge
density in the free field theory, because to the leading order 
$\partial{\overline{\partial}}x=\partial\overline{\partial}\vartheta=0$.
But local conserved charges
in a free field theory are all quadratic in the free fields, there are
no local conserved charges cubic or of higher order. Therefore, $d\Phi=0$
implies that $\Phi$ is an exact form.

These arguments would
not work if $\Phi$ was quadratic in the elementary fields. For example
for $\delta[x,dx]=d[\delta x, x]+\ldots$ where dots are higher order
terms, because the leading term in $\delta x=\xi$ is constant. 
But $[x,dx]$ is not a total derivative, not even a closed form.} 
$\Phi=dG$, and therefore the log divergence can be gauged away by a $z$-dependent
gauge transformation.

This proves the absence of the 1-loop logarithmic divergences of the form 
$F(x,\vartheta)dx$ and $F(x,\vartheta)d\vartheta$. Another possibility
would be the logarithmic divergences of the form $F(x,\theta)N_{\pm}$.
But in Section \ref{sec:DivergenceNPlus} we have shown that there are no
log divergences proportional to $N_{\pm}$ without $x$ and $\theta$.
Therefore the lowest order terms in the near flat space
expansion would contain at least one $x$ or $\theta$. Such terms cannot be invariant
under global shifts, and therefore should cancel.  

It is not surprising that global symmetries relate the log divergences at finite
$x$ to the log divergences at $x=0$, and therefore it is enough to prove that
there are no divergences proportional to $\partial_{\pm}x$ and $\partial_{\pm}\vartheta$.
It should be possible to reach the same conclusion using the background field
method \cite{Puletti:2006vb}.

\subsection{Singularity in the product $J_{2+}J_{2+}$ and global shifts}
\label{sec:SingularityJ2J2AndGlobalShifts}
We have $J_{2+}=-\partial_+x +\ldots$ and 
$J_{2+}^{\mu}(w_L) J_{2+}^{\nu}(w_R)=
-{1\over R^2}{C^{\mu\nu}\over (w_L-w_R)^2}+\ldots$. Let us consider the
variation of $J_{2+}$ under the global shift. According to 
(\ref{CompensatingGaugeTransformation}) we have 
$\delta_{\xi}J_{2+}={1\over 2}[[x,\xi],J_{2+}]$. Consider the
variation of the product:
\begin{eqnarray}
&& \delta_{\xi}(J_{2+}^{\mu}(w_L) J_{2+}^{\nu}(w_R)) = 
{1\over 2}[[x,\xi],\partial_+x]^{\mu}(w_L) \partial_+x^{\nu}(w_R)
+{1\over 2}\partial_+x^{\mu}(w_L)  [[x,\xi],\partial_+x]^{\nu}(w_R)
=\nonumber
\\
&& =-{1\over 2}{1\over (w_L-w_R)^2}[[x(w_L),\xi],t^{2\nu}]^{\mu}
+{1\over 2}{1\over (w_L-w_R)}[[t^{2\nu},\xi],\partial_+x]^{\mu}+
\ldots+(w_L\leftrightarrow w_R , \mu \leftrightarrow \nu)
=\nonumber 
\\
&& ={1\over 2} {\bar{w}_L-\bar{w}_R\over (w_L-w_R)^2}
\mbox{str}([\partial_-x,\xi][t^{2\mu},t^{2\nu}])
+\ldots
\label{VariationOfTheProduct}
\end{eqnarray}
where $t^{2\mu}=t^2_{\nu}C^{\nu\mu}$. 
On the other hand from Sections \ref{sec:DivXXzMinus4} and 
\ref{sec:DivXXzPlus4} and the ``symmetry" (\ref{MetaSymmetry})  we know that
\[
J_{2+}^{\mu}(w_L) J_{2+}^{\nu}(w_R)\;[t^2_{\mu},t^2_{\nu}]=
-{1\over 4} {\bar{w}_L - \bar{w}_R\over (w_L - w_R)^2}
[\partial_- x , C_{\bar{2}}.x] +\ldots
\]
This formula is in agreement with (\ref{VariationOfTheProduct}) because 
for any $\xi^2\in {\bf g}_2$ and $\eta^2\in {\bf g}_2$ we have:
\[
\mbox{str}([\xi^2,\eta^2] [t^{2\mu},t^{2\nu}])[t^2_{\mu},t^2_{\nu}]
 = -{1\over 2} [\xi^2,C_{\bar{2}}.\eta^2] 
\]

\section{Infinite line}
\label{sec:InfiniteLine}
\subsection{Transfer matrix on the infinite line}
We will define the transfer matrix on the infinite line
as the limit:
\begin{equation}
\lim_{\begin{array}{c} \tau_l\to +\infty \\ 
			\tau_r\to -\infty \end{array}}
(\Omega_{\tau_r}^{\tau_l}(z=1))^{-1} \; \Omega_{\tau_r}^{\tau_l}(z) \,.
\end{equation}
This can be expressed through the ``small case currents":
\begin{eqnarray}
& P\exp\int_{-\infty}^{\infty} 
&\left[ 
 \left((1-z^{-1})j_{3+} + (1-z^{-2})j_{2+} + (1-z^{-3}) j_{1+}
+(1-z^{-4})j_{0+} \right)d\tau^+ + \right.
\nonumber\\
&&
\left. \left((1-z)j_{1-} + (1-z^2)j_{2-} + (1-z^3) j_{3-}
+(1-z^4)j_{0-} \right)d\tau^- 
\right] \,.
\end{eqnarray}
The definition of the transfer matrix on the infinite
line is such that the power of $z$ does not correlate
with the $\mathbb{Z}_4$ grading. This is because we divided
by $\Omega(z=1)$.

\subsection{Global symmetry charge}
It is useful to check our formalism by showing that the
global Lorentz charge is finite. We will verify
that there are no divergences proportional to $\partial_+ x$, $[\partial_+ x ,x]$ and to  $N_+$. Consider the charge
corresponding to  boosts and rotations around the 
point $x=0$. This charge can be computed by expanding
the transfer matrix on the infinite line in
$z=1+\zeta$ to the first order in $\zeta$:
\begin{equation}
q_{global}= \int \; * g^{-1}_{z=1} 
\left[
\left.{\partial\over\partial z}\right|_{z=1} J(z)  
\right]
g_{z=1}\,.
\end{equation}

Note that we will omit the powers of $1/R$, since these are obvious (each $x$ and $\vartheta$ comes with one power of $1/R$) and would only clutter the formulas.

\vspace{10pt}

\noindent \underline{\em Divergences proportional to $\partial_+ x $}
The terms responsible for the logarithmic divergences of the
global current proportional to $j_+$ are:
\begin{eqnarray}
&&	\left( 1 - {1\over z^2}\right) (g^{-1} J_{2+} g)_{\bar{2}} +
	\left( 1 - {1\over z}\right)   
	\left(	[\vartheta_L, \partial_+\vartheta_L] -
		[\vartheta_L, [\partial_+ x , \vartheta_R] ] 
	\right) +
\nonumber \\
&&	\left( 1 - {1\over z^3}\right) 
	\left(	[\vartheta_R, \partial_+\vartheta_R] -
		[\vartheta_R, [\partial_+ x , \vartheta_L] ] 
	\right) + \ldots \,.
\nonumber
\end{eqnarray}
Therefore the $\partial_+x$-part of the log divergence of $j_+$ is
 the same as the log divergence of:
\begin{eqnarray}
&&	2 J_{2+} - \left( [ x , [ x, \partial_+ x] ]
		+ [ \vartheta_L , [\vartheta_R , \partial_+ x ] ]
		+ [ \vartheta_R , [\vartheta_L , \partial_+ x ] ]
			\right) 
\nonumber \\
&&
+ [\vartheta_L , (\partial_+ \vartheta_L - [\partial_+ x , \vartheta_R])]
+ 3 [\vartheta_R , (\partial_+ \vartheta_R - [\partial_+ x , \vartheta_L])]
+\ldots \,.
\end{eqnarray}
Notice that $[\vartheta_L , \partial_+ \vartheta_L]$ does not contribute
to the log divergence, while $3 [\vartheta_R , \partial_+ \vartheta_R]$
contributes the same amount as 
$3 [\vartheta_R , [\partial_+x , \vartheta_L ] ]$.
The contribution of $[ \vartheta_L , [\vartheta_R , \partial_+ x ] ]$
is minus the contribution of $[ x , [ x, \partial_+ x] ]$.
Therefore at the order $R^{-3}$ the $\partial_+ x$-piece of the
log divergence in $j_+$ is the
same as the $\partial_+ x$-piece of the log divergence in $2J_{2+}$.
But we have seen in Section \ref{sec:FieldRenormalization}
that $2J_{2+}$ to the order $R^{-3}$ is finite.
This shows that the log divergence of $j_+$ proportional to 
$\partial_+ x$ is zero at the order $R^{-3}$.

\vspace{10pt}

\noindent \underline{\em Divergences proportional to $[\partial_+ x, x]$}
We will split the calculation into two parts, first identifying the
contribution of bosons, and then the contribution of fermions.
The contribution of bosons comes from
\begin{equation}
-2\mbox{Ad}(e^{-x}).J_{2+}=
2\mbox{Ad}(e^{-x}).
\left(\partial_+ x^{ren} + 
{1\over 6} :[x^{ren}, [x^{ren}, \partial_+x^{ren}]]:\right)\,.
\end{equation}
We have taken into account that the coefficient of $\partial_+ x$ in
$J_{2+}$ is not renormalized, and therefore we should skip the
contractions of $x^{ren}$ with $x^{ren}$ in
$[x^{ren}, [x^{ren}, \partial_+x^{ren}]]$, as denoted by the double
dots. Therefore, the log divergence is that of the expression:
\begin{equation}
-2 [ x, \partial_+ x^{ren} ] - {1\over 3} [x, [x, [x, \partial_+ x]]]
-{1\over 3} [x, :[x, [x, \partial_+ x]]:]\,.
\end{equation}
The relation between $x$ and $x^{ren}$ is given by 
Eq. (\ref{RenormalizationOfX}). Let us take only the term 
generated by the bosons:
\begin{eqnarray}
x& =& x^{ren}+{1\over 6} {1\over R^2}\log\epsilon^2\;
[t^2_{\mu} , [t^2_{\mu} , x^{ren}] ]+\\
&&+\mbox{terms generated by fermions} \,.
\end{eqnarray}
The result is:
\begin{equation}\label{IntermediateResultBosonicXX}
2[\partial_+x^{ren}, x^{ren}]-{1\over 2} \log\epsilon^2 
[\partial_+ x, C_{\bar{2}}.x] \,.
\end{equation}
But the expression $2[\partial_+x^{ren}, x^{ren}]$ itself
has an internal log divergence, because of the interaction vertex
$-{1\over 6}[\partial_+x, x][\partial_-x , x]$ in the action.
The log divergence of $2[\partial_+x^{ren}, x^{ren}]$ is the same as
of the expression ${2\over 3} [x, :[x,[x,\partial_+ x]]:]$ and equals
to ${1\over 2} \log\epsilon^2 [\partial_+ x, C_{\bar{2}}.x]$.
This cancels $-{1\over 2} \log\epsilon^2 
[\partial_+ x, C_{\bar{2}}.x]$ in (\ref{IntermediateResultBosonicXX})
and gives {\em the total of zero} from bosons.

Now let us evaluate the contribution of fermions. 
One source of contribution is:
\begin{equation}\label{FermionicContributionFromJ2}
-2\mbox{Ad}(e^{-x}e^{-\vartheta}).J_{2+}\,.
\end{equation}
The relevant terms are 
\begin{equation}\label{ExpandingExponential}
2[\partial_+ x^{ren}, x]
-[x, [\vartheta_L, [\vartheta_R, \partial_+ x]]]
-[x, [\vartheta_R, [\vartheta_L, \partial_+ x]]]\,.
\end{equation}
Here we want
to pick the fermionic contribution to the renormalization of $x$:
\begin{eqnarray}
x & = & x^{ren} + 
{1\over 2} {1\over R^2}\log\epsilon^2\; C^{\alpha\dot{\beta}}
\{ t^3_{\alpha} , [ t^1_{\dot{\beta}}, x^{ren}] \} + \nonumber\\
&&+\mbox{contribution of bosons} \,.
\end{eqnarray}
Substitution of this formula into (\ref{ExpandingExponential})
gives the total contribution from (\ref{FermionicContributionFromJ2})
equal to 
\begin{equation}\label{FermionicToXXFromJ2}
\log\epsilon^2
[\partial_+x, C_{\bar{2}}.x]\,.
\end{equation}
The other source of fermionic contributions is
\begin{equation}
-\mbox{Ad}(e^{-x}e^{-\vartheta}).(3J_{1+}+J_{3+})\,.
\end{equation}
The relevant terms are:
\begin{eqnarray}
&	\mbox{Ad}(e^{-x}e^{-\vartheta}). & \left(
3\partial_+\vartheta_R + 
3[\vartheta_L,\partial_+x] +
{3\over 2}[\vartheta_R,[x,\partial_+x]] +\right. \\ &&\left.\;\;\;
\partial_+\vartheta_L + [\vartheta_R,\partial_+x] +
{1\over 2}[\vartheta_L,[x,\partial_+x]]
\right)\,.
\end{eqnarray}
The log divergence of this is the same as of the expression
\[
3[\partial_+\vartheta_R, \vartheta_L]  + [\partial_+\vartheta_L, \vartheta_R]
+[\partial_+x, [\vartheta_R,[\vartheta_L,x]]]\,,
\]
and is equal to 
\begin{equation}\label{FermionicToXXFromJodd}
-\log\epsilon^2 [\partial_+x, C_{\bar{2}}.x]\,.
\end{equation}
Here we have taken into account that $4[\partial_+\vartheta_R, \vartheta_L]$
mixes into $[\partial_+ x, x]$ because of the following interaction 
vertices in the action:
\[
- {1\over 8}  [\vartheta_L, \partial_- \vartheta_R] [x, \partial_+ x]
- {3\over 8}  [\vartheta_R, \partial_- \vartheta_L] [x, \partial_+ x]\,.
\]
We see that the total contribution from fermions
(\ref{FermionicToXXFromJ2}) + (\ref{FermionicToXXFromJodd})
is also zero, and therefore the Lorentz current is not renormalized.
\vspace{10pt}

\noindent \underline{\em Divergences proportional to $N_+$}
The terms responsible for the logarithmic divergence
proportional to $N_+$ are:
\begin{eqnarray}
&& \left(1-{1\over z^4}\right)
N_+ + {1\over 2} {1\over R^2} 
\left(1-{1\over z^4}\right)
([x , [x , N_+]] + [\vartheta_L , [\vartheta_R , N_+]]
+[\vartheta_R , [\vartheta_L , N_+]])
+\nonumber \\ &&+
{1\over R^2} \left({1\over z^2}-1\right) 
[\partial_+ x , x]  +
 {1\over R^2} \left({1\over z}-1\right) 
[\partial_+ \vartheta_L , \vartheta_R] + 
{1\over R^2} \left({1\over z^3}-1\right) 
[\partial_+ \vartheta_R , \vartheta_L] 
+\ldots \,.
\nonumber
\end{eqnarray}
When we expand this in powers of $\zeta$, the linear term is:
\begin{eqnarray}
4 N_+ &+& {2\over R^2}
\left([x , [x , N_+]] + [\vartheta_L , [\vartheta_R , N_+]]
+[\vartheta_R , [\vartheta_L , N_+]]\right)
-\label{NPlusInGlobalCharge} \\ 
&-&{2\over R^2}
\left( {1\over 2}[\partial_+ \vartheta_L , \vartheta_R] +
[\partial_+ x , x]  +
{3\over 2} [\partial_+ \vartheta_R , \vartheta_L] \right)
+\ldots \,.
\nonumber
\end{eqnarray}
It follows from Eq. (\ref{CasimirOnVector}) that the log divergence
of the second line is equal to the log divergence of:
\begin{equation}
-{2\over R^2}
\left(
	[\partial_+ \vartheta_L , \vartheta_R] +
	[\partial_+ x , x]  +
	[\partial_+ \vartheta_R , \vartheta_L] 
\right) = -4 J_{0+} \,.
\end{equation}
Eq. (\ref{LeftInternalJ0ToN}) shows that the mixing of $-4 J_{0+}$
into $N_+$ cancels the log divergence of the first line in
(\ref{NPlusInGlobalCharge}).
This shows that the log divergence of $j_+$ proportional to 
$N_+$ is zero at the order $R^{-4}$.

Of course, finiteness of $j_+$ is guaranteed by the quantum worldsheet theory
being invariant under the global symmetries.

\vspace{10pt}

\section{Summary and Conclusions}
\label{sec:SummaryAndConclusions}
We have shown that the logarithmic divergences of the transfer
matrix in the pure spinor superstring in $AdS_5\times S^5$ vanish
at the one loop level. 
The Lax operator in the pure spinor string in $AdS_5\times S^5$,
although it looks somewhat cumbersome, seems to work beautifully in the
quantum theory. It is reasonable to conjecture that the path ordered
exponential of the Lax connection defines a sensible quantum transfer
matrix. Notice that in the previously studied examples of massive
integrable systems the transfer matrix had logarithmic divergences,
while in our case it seems that only the linear divergences are
present. 

It would be very interesting to investigate the quantum commutation
relations of the components of this transfer matrix, and see
if they could be encoded in the form of the RTT relations. In principle
this could be done in perturbation theory, in the near-flat space limit.


\subsection*{Acknowledgments}

We thank N. Berkovits and A. Tseytlin for discussions and comments on the draft.
The research of AM was supported by the Sherman Fairchild 
Fellowship and in part
by the RFBR Grant No.  03-02-17373 and in part by the 
Russian Grant for the support of the scientific schools
NSh-1999.2003.2. 
The research of SSN was supported by a John A. McCone Postdoctoral Fellowship of Caltech. 


\newpage

\setcounter{section}{0} 

\appendix{The algebra $\mathfrak{psu}(2,2|4)$}
\label{app:Algebra}

\subsection{Structure constants and invariant tensor}

Consider the quadratic Casimir operator:
\begin{equation}
C=C^{\dot{\alpha}\alpha} 
(	 t_{\dot{\alpha}}^1\otimes t_{\alpha}^3 
	-t_{\alpha}^3\otimes t_{\dot{\alpha}}^1	)
 +C^{\mu\nu} t_{\mu}^2\otimes t_{\nu}^2 
 +C^{[\mu_1\mu_2][\nu_1\nu_2]}
t_{[\mu_1\mu_2]}^0\otimes t_{[\nu_1\nu_2]}^0 \,.
\end{equation}
We also define $C_0$:
\begin{equation}
C_0= C^{[\mu_1\mu_2][\nu_1\nu_2]}
t_{[\mu_1\mu_2]}^0\otimes t_{[\nu_1\nu_2]}^0 \,.
\end{equation}
We define $C^{\alpha\dot{\alpha}}$ and $C$ with lower indices as follows:
\begin{equation}
C^{\alpha\dot{\alpha}}=-C^{\dot{\alpha}\alpha} \; ,\;\;
C^{\alpha\dot{\beta}} C_{\dot{\beta}\gamma} = \delta_{\gamma}^{\alpha} \,.
\end{equation}
The structure constants with upper indices are defined as:
\begin{equation}
\fduu{\dot{\gamma}}{\alpha}{\mu}=
\fddu{\dot{\gamma}}{\dot{\alpha}}{\mu}
C^{\dot{\alpha}\alpha}\; , \;\;
\fduu{\gamma}{\dot{\alpha}}{\mu}=
\fddu{\gamma}{\alpha}{\mu}
C^{\alpha\dot{\alpha}} \,.
\end{equation}
Similarly, the vector indices are raised by $C^{\mu\nu}$. Notice that $C^{\mu\nu}$
is a symmetric tensor. If we identify $t_{\mu}^2$ with the Killing vectors
on $AdS_5\times S^5$ then $C^{\mu\nu}=g^{\mu\nu}$ should be identified
with the metric.
We have \[
\fduu{\dot{\gamma}}{\alpha}{\mu}=-\fduu{\dot{\gamma}}{\mu}{\alpha} \,.
\]
We will normalize the supertrace so that:
\begin{equation}
\mbox{str}(t_{\mu}^2t_{\nu}^2)=g_{\mu\nu}\; , \;\;
\mbox{str}(t_{\dot{\alpha}}^1 t_{\beta}^3)=C_{\dot{\alpha}\beta}\; , \;\;
\mbox{str}(t_{\beta}^3 t_{\dot{\alpha}}^1)= C_{\beta\dot{\alpha}} \,.
\end{equation}

\subsection{Matrix realization}

The algebra $\mathfrak{sl}(4|4)$ can be realized by the $(4|4)\times (4|4)$-matrices 
of the form:
\begin{equation}
M = \left( \begin{array}{cc} A & X \\
Y & B \end{array} \right) \,.
\end{equation}
The $\mathbb{Z}_4$ automorphism is $M\mapsto \Omega M$ where:
\begin{equation}
\Omega M = 
\left(
\begin{array}{cc}
J A^t J  &  -J Y^t J \\
J X^t J  &   J B^t J 
\end{array}
\right) \,,
\end{equation}
where $J$ is an antisymmetric matrix:
\begin{equation}
J=\left(\begin{array}{cccc}  0	&   0	&  -1	&  0	\\
                             0  &   0	&   0	&  1	\\
			     1  &   0   &   0   &  0    \\
			     0  &  -1	&   0	&  0	
	\end{array}\right) \,.
\end{equation}
Notice that $\Omega^4 M = M$. To get $\su(2,2|4)$ from $\mathfrak{sl}(4|4)$ we
impose the reality condition $M^{\dagger} = -M$, where
\begin{equation}
M^{\dagger} = 
\left(
\begin{array}{cc}
\Sigma A^{\dagger} \Sigma & -i\Sigma Y^{\dagger} \\
- i X^{\dagger} \Sigma &    B^{\dagger} 
\end{array}
\right)\,.
\end{equation}
The subspaces $\mathfrak{g}_{\bar{a}}$ are defined as eigenspaces of $\Omega$:
\begin{equation}
\Omega \xi = i^a \xi \;\;\;\mbox{for}\;\; \xi\in {\mathfrak{g}}_{\bar{a}} \,.
\end{equation}
Notice that ${\mathfrak{g}}_{\bar{a}}\cap \su (2,2|4)$, $\bar{a}\in \{0,1,2,3\}$ are real
subspaces of $\psu (2,2|4)$. (And therefore $\Omega$ does not really act
on $\su (2,2|4)$ but only on $\mathfrak{sl}(4|4)$.)
The AdS superalgebra $\psu(2,2|4)$ is a factoralgebra of $\su (2,2|4)$ 
by the center, which is generated by the unit matrix. 

The invariant bilinear form on the superalgebra $\psu(2,2|4)$ can be
defined using the supertrace in the {\em fundamental} representation:
\begin{equation}
C(\xi,\eta) = \mbox{str}\;\xi\eta \,.
\end{equation}

\subsection{Some algebraic identities}
Here we collect some useful algebraic identities.
Notice that:
\begin{equation}\label{QuadraticSerre}
C^{\alpha\dot{\alpha}} \{ t^3_{\alpha} , t^1_{\dot{\alpha}} \} = 0
\end{equation}

\noindent {\bf First identity.}
If X is a spinor (an element of ${\mathfrak{g}}_{\bar{1}}$ or ${\mathfrak{g}}_{\bar{3}}$)
then 
\begin{eqnarray}
&&	C^{\dot{\alpha}\alpha}[t^1_{\dot{\alpha}} , \{ t^3_{\alpha},X \} ]
	=C^{\alpha\dot{\alpha}}[t^3_{\alpha} , \{t^1_{\dot{\alpha}},X\} ]
	=C^{\mu\nu} [t^2_{\mu},  [t^2_{\nu}, X] ]=
\label{NoSpinorDiv}\\
&&	=C^{[\mu_1\nu_1][\mu_2\nu_2]} 
	[ t^0_{[\mu_1\nu_1]} , [ t^0_{[\mu_2\nu_2]} , X ] ] = 0
\qquad \qquad  \mbox{if}\;\; X\in {\mathfrak{g}}_{\bar{1}}+{\mathfrak{g}}_{\bar{3}} \,.
\nonumber
\end{eqnarray}
The total adjoint Casimir of $\psu(2,2|4)$ is zero. Therefore
it is enough to prove that these four expressions in (\ref{NoSpinorDiv})
are equal to  each other. Notice that
\begin{equation}
C^{\alpha\dot{\alpha}}\{ t^3_{\alpha} , t^1_{\dot{\alpha}} \} = 0 \,,
\end{equation}
because this would be in the center of ${\mathfrak{g}}_0$, but the center
of ${\mathfrak{g}}_0$ is trivial.
This implies the equality of the first two expressions in (\ref{NoSpinorDiv}).
The other equalities can be demonstrated as follows: 
\begin{eqnarray}
&&	C^{\alpha\dot{\alpha}} 
[t_{\alpha}^3 , \{t_{\dot{\alpha}}^1 , t_{\beta}^3 \} ] =
-[t_{\alpha}^3 , 
	\fduu{\!\!\!\!\!\beta}{\;\;\alpha}{[\mu\nu]} t_{[\mu\nu]}^0]=
\nonumber \\
&&	=C^{[\mu_1\nu_1][\mu_2\nu_2]} 
	[ t^0_{[\mu_1\nu_1]} , [ t^0_{[\mu_2\nu_2]} , t_{\beta}^3 ] ]
\\[7pt]
&&	C^{\dot{\alpha}\alpha} 
	[t_{\dot{\alpha}}^1 , \{ t^3_{\alpha} , t^3_{\beta} \} ] 
	=-[t_{\dot{\alpha}}^1, \fduu{\beta}{\dot{\alpha}}{\mu} t_{\mu}^2] =
\nonumber\\
&&	=C^{\mu\nu} [t_{\mu}^2 , [t_{\nu}^2, t_{\beta}^3] ] \,.
\end{eqnarray}

\vspace{10pt}
\noindent {\bf Second identity.}
If $X\in {\mathfrak{g}}_{\bar{2}}$ then 
\begin{eqnarray}
&&	C^{\dot{\alpha}\alpha}[t^1_{\dot{\alpha}} , \{ t^3_{\alpha},X \} ]
	=C^{\alpha\dot{\alpha}}[t^3_{\alpha} , \{t^1_{\dot{\alpha}},X\} ]
	=-C^{\mu\nu} [t^2_{\mu},  [t^2_{\nu}, X] ]=
\label{CasimirOnVector}\\
&&	=-C^{[\mu_1\nu_1][\mu_2\nu_2]} 
	[ t^0_{[\mu_1\nu_1]} , [ t^0_{[\mu_2\nu_2]} , X ] ] 
\qquad \qquad  \mbox{if}\;\; X\in {\mathfrak{g}}_{\bar{2}} \,.
\nonumber
\end{eqnarray}
Notice that in this case, when $X\in {\mathfrak{g}}_{\bar{2}}$, these three
expressions are equal to each other, but not zero.

\vspace{10pt}
\noindent{\bf Third identity.}
We define $C_0$ as follows:
\begin{equation}
C_0 = C^{[\mu_1\nu_1][\mu_2\nu_2]} 
t^0_{[\mu_1\nu_1]}\otimes t^0_{[\mu_2\nu_2]} \,.
\end{equation}
If $X\in {\mathfrak{g}}_{\bar{0}}$ then
\begin{equation}
C_0.X=C^{[\mu_1\nu_1][\mu_2\nu_2]} 
	[ t^0_{[\mu_1\nu_1]} , [ t^0_{[\mu_2\nu_2]} , X ] ]
=\left\{
\begin{array}{rl}
-6 X \;\;\;\;& \mbox{if}\;\; X\in \mathfrak{so}(1,4) \\
6 X \;\;\;\; & \mbox{if}\;\; X\in \mathfrak{so}(5)
\end{array}\right. \,.
\end{equation}
{\bf Casimir identities.}
If $\xi_{\bar{2}}\in \mathfrak{g}_{\bar{2}}$ then 
the adjoint Casimir satisfies:
\begin{eqnarray}
	(C_{\bar{0}}+C_{\bar{2}}).[\xi_{\bar{2}},\eta_{\bar{2}}] & = &
	[\xi_{\bar{2}}, (C_{\bar{0}}+C_{\bar{2}}).\eta_{\bar{2}}]
\\
	C_{\bar{0}}.\xi_{\bar{2}} & = & C_{\bar{2}}.\xi_{\bar{2}}
\\
	C_{\bar{2}}.[\xi_{\bar{2}},\eta_{\bar{2}}] & = &
	{1\over 2} [\xi_{\bar{2}} , C_{\bar{2}}.\eta_{\bar{2}}]
\\
	C_{\bar{0}}.[\xi_{\bar{2}},\eta_{\bar{2}}] & = &
	{3\over 2} [\xi_{\bar{2}} , C_{\bar{2}}.\eta_{\bar{2}}] \,.
\end{eqnarray}
We will also introduce
\begin{equation}
C_{odd}=C_{\bar{1}}+C_{\bar{3}}\;\;, \;\;\;
C_{even}=C_{\bar{0}}+C_{\bar{2}} \,.
\end{equation}
Notice that for $\xi_{\bar{2}}\in \mathfrak{g}_{\bar{2}}$:
\begin{equation}\label{CoddAndC2}
C_{odd}.\xi_{\bar{2}}=-2C_{\bar{2}}.\xi_{\bar{2}} \,.
\end{equation}
{\bf Additional identities.}
\begin{eqnarray}
&&	C^{\alpha\dot{\alpha}} 
(	t^3_{\alpha} t^2_{\mu} t^1_{\dot{\alpha}} +
	t^1_{\dot{\alpha}} t^2_{\mu} t^3_{\alpha} 
)=0
\nonumber
\\
&&	
	[ t_{\mu}^2 \; , \; C^{\alpha\dot{\alpha}} 
	  t^3_{\alpha} t^1_{\dot{\alpha}} ] = 0 \,.
\label{MoreIdentities}
\end{eqnarray}


\appendix{Contour-split regularization and linear divergences}
\label{app:LinearDivergences}

\subsubsection{Regularization by splitting along the contour}

We treat $x$, $\vartheta$, $\lambda$ and $w$
 as elementary fields. The
capital currents $J_{\pm}$ are composite operators constructed
from these elementary fields. When calculating the Wilson loop,
we assume that the composite operators are regularized
by splitting {\em along the contour}. For example, the expression
$[ x , [ x , \partial_+ x] ]$ will be understood as follows:
\begin{equation}
 x(\tau+2\epsilon) x(\tau+\epsilon) \partial_+x(\tau) -
2x(\tau+2\epsilon) \partial_+x(\tau+\epsilon) x(\tau) +
 \partial_+x(\tau+2\epsilon) x(\tau+\epsilon) x(\tau) \,.
\end{equation}

\vspace{10pt}

\begin{centering}
	\hfill
	\epsfxsize=3in
      \epsffile{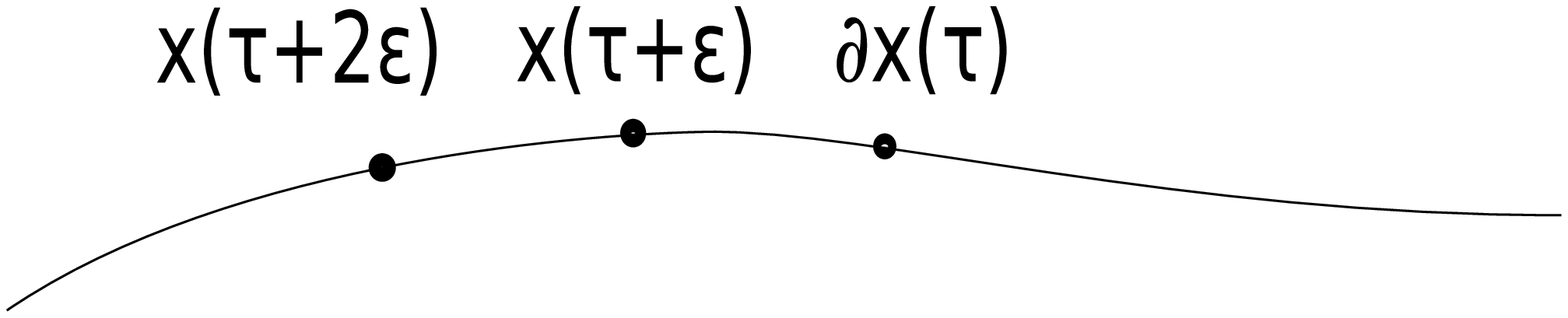}
      \hfill
\end{centering}

\vspace{10pt}

Here $\tau+\epsilon$ denotes the shift of the
point $\tau$ by the amount $\epsilon$ along the contour.
This regularization depends on the parametrization of the contour.
The good thing about this regularization is that it preserves
the property that the transfer matrix is a total derivative when
$z=1$. 
This regularization can be understood as introducing the
``dot-product":
\begin{equation}
(\phi_1 \cdot \phi_2)(\tau) = \phi_1(\tau) \phi_2(\tau-\epsilon) \,.
\end{equation}
The integrated dot-product is associative:
\begin{equation}
\int d\tau\; \phi_1\cdot (\phi_2 \cdot \phi_3) =
\int d\tau\; (\phi_1\cdot \phi_2) \cdot \phi_3 \,.
\end{equation}
This property implies that the regularized transfer matrix is a total 
derivative when $z=1$, just like it was in the classical theory:
\begin{equation}
P\exp \left(-\int_{\tau_1}^{\tau_2} J^{[z=1]}\right) = 
g(\tau_1) g(\tau_2)^{-1} \,.
\end{equation}
Here $g(\tau)$ is, schematically, 
$e^{x(\tau)}=1+x+{x\cdot x\over 2} + {x\cdot x\cdot x\over 6}+\ldots$.
We also introduce the split commutator $\lbr , \rbr$:
\begin{equation}
\lbr \phi_1 , \phi_2 \rbr =
\phi_1\cdot\phi_2 - \phi_2\cdot \phi_1 \,.
\end{equation}


\subsubsection{Example: Wilson line in the $O(2)$ nonlinear sigma-model}

The $O(2)$ NLSM is a free field theory, a single boson $\phi$
with the action $\int d^2\tau \partial_+ \phi \partial_- \phi$.
We want to define the Wilson line:
\begin{equation}
P\exp \int_0^L \left[ {1\over z^2} \partial_+\phi d\tau^+ +
z^2 \partial_-\phi d\tau^- \right]
\end{equation}
This is very easy to calculate, if we allow ourselves to split
$\phi$ into holomorphic and antiholomorphic part:
\begin{equation}
\phi(\tau^+,\tau^-)= \phi_{h}(\tau^+) + \phi_{a}(\tau^-) \,.
\end{equation}
The transfer matrix is equal to:
\begin{equation}
e^{{1\over z^2}\phi_{h}(L) + z^2 \phi_{a}(L)}
e^{-{1\over z^2}\phi_{h}(0) - z^2 \phi_{a}(0)} \,.
\end{equation}
But we want to insist on using our contour-split regularization.
Therefore we should calculate the transfer matrix as
$
P\exp \left( -\int J^{[z]} \right) \,,
$
where
\begin{equation}
J^{[z]}=-d \exp \left[ {1\over R}
\left({1\over z^2} \phi_{h} + z^2 \phi_{a}\right)\right]
\exp\left[-{1\over R}\left({1\over z^2} \phi_{h} + z^2 \phi_{a}\right)
\right] \,.
\end{equation}
We should expand in powers of $1\over R$ and use our dot-product
for $\phi$, for example replace $\phi^2(\tau)$ with 
$\phi(\tau)\phi(\tau-\epsilon)$. 
One of the terms we get is:
\begin{equation}
{1\over R^2}{1\over z^4} {1\over 2}\lbr d\phi_{h} , \phi_{h} \rbr =
-{1\over R^2}{1\over z^4}  {1\over \epsilon^+} \slot \slot d\tau^+  \,.
\end{equation}
Here $\slot$ is a ``placeholder", for example
$\phi \slot \phi = \phi(\tau+2\epsilon)\phi(\tau)$ and
$\phi \slot\slot \phi = \phi(\tau+3\epsilon)\phi(\tau)$; 
the placeholder becomes important when we study the collisions.
A more complicated example is:
\begin{eqnarray}
&&	{1\over R^3}{1\over z^6} {1\over 6}
	\lbr \lbr d\phi_{h} , \phi_{h} \rbr , \phi_{h} \rbr = 
\\
&&	={1\over R^3}{1\over z^6} {1\over 6} d\tau^+ 
\left({3\over\epsilon^+} \lbr \phi_{h} , \slot\slot \; \rbr
-\log\epsilon \lbr\lbr \partial_+\phi , \slot \; \rbr , \slot \; \rbr
+ 2\log 2 \slot \partial_+ \phi \slot
\right) \,.
\end{eqnarray}
Notice that this expression is almost expressed in terms of the
derivatives of $\phi$, except for the term 
${3\over\epsilon^+} \lbr \phi_{h} , \slot\slot \; \rbr$.
This term can be expressed through the derivatives of $\phi$ in the bulk;
in the boundary terms $\phi_{h}$ enters either through explicit
contractions, or as a derivative.
In fact, we should probably think of any commutator with the placeholder
as a total derivative, because it plays a role only in the boundary
terms.


\subsubsection{Linear divergences}
Since $J$-currents are built on both $x$ and $\partial x$, 
there are linear divergences in them. The coefficient of the 
linear divergence is either a c-number or a function of $x$ 
(but no derivatives of $x$).
For example, $J_{0+}$ has a term ${1\over 2}[\partial_+ x, x]$,
which leads to the linear divergence:
\begin{equation}
{1\over 2}\lbr \partial_+ x, x \rbr = -{1\over \epsilon^+} 
C^{\mu\nu} t^2_{\mu} t^2_{\nu} +\ldots \,.
\end{equation}
Therefore we have to add the conterterms to the currents, to make the 
transfer finite, for example:
\begin{equation}
J_{0+}^{with\;c.t.}=J_{0+} + {1\over \epsilon^+} 
\left( 
C^{\mu\nu} t^2_{\mu} t^2_{\nu} +
C^{\alpha\dot{\alpha}} t^3_{\alpha} t^1_{\dot{\alpha}} +
C^{\dot{\alpha}\alpha} t^1_{\dot{\alpha}} t^3_{\alpha}
\right) \,.
\end{equation} 
Notice that these counterterms
do not belong to the Lie algebra.
Similarly, $J_{1+}$, $J_{2+}$ and $J_{3+}$ have linear divergences,
as composite operators. In the expansion 
of $J_{2+}$, the terms responsible for the linear divergence are:
\begin{equation}
J_{2}  =  -d x -
{1\over 6}  \lbr x, \lbr x , d x \rbr \rbr + \ldots \,.
\end{equation}
This leads to the following linear divergence in $J_{2+}$:
$
-{1\over 2}{1\over \epsilon^+} 
\lbr  x , C^{\mu\nu} t^2_{\mu} t^2_{\nu}  \rbr \,.
$.
But this linear divergence in fact cancels with the linear divergence
arizing in the $J_{0+}J_{2+}$ collision.
Therefore there is no linear counterterm to $J_{2+}$.

The double collisions  $J_{2+} J_{2+}$ and $J_{1+}J_{3+}$ give nonzero
linear divergences which should be cancelled
by the counterterm proportional to $z^{-4}$. There is no classical $J_{4+}$
current, but we need to introduce such a counterterm to cancel the
linear divergence:
\begin{equation}
{1\over z^4} J_{4+}^{c.t.} =
-{1\over z^4} {1\over R^2} {1\over \epsilon^+} 
\left( 
C^{\mu\nu} t^2_{\mu} t^2_{\nu} +
C^{\alpha\dot{\alpha}} t^3_{\alpha} t^1_{\dot{\alpha}} +
C^{\dot{\alpha}\alpha} t^1_{\dot{\alpha}} t^3_{\alpha}
\right) \,.
\end{equation}
There is a linear divergence in the collisions $J_{0+}J_{1+}$
and $J_{3+}J_{2+}$,
which cancels with the internal linear divergence of $J_{1+}$.
Similarly, the linear divergence in the collisions
$J_{0+}J_{3+}$ and $J_{1+}J_{2+}$ 
cancels with the internal linear divergence of $J_{3+}$.


\newpage

\bibliographystyle{JHEP} \renewcommand{\refname}{Bibliography}
\addcontentsline{toc}{section}{Bibliography}


\providecommand{\href}[2]{#2}\begingroup\raggedright\endgroup

\end{document}